\documentclass[aps,prb,reprint,superscriptaddress,a4paper]{revtex4-1}

\usepackage{graphicx}
\usepackage{dcolumn}

\begin{document}

\title{Fluctuations in the electron system of a superconductor exposed to a photon flux} 

\author{P.J. de Visser}
\email{p.j.devisser@tudelft.nl}

\affiliation{SRON National Institute for Space Research, Sorbonnelaan 2, 3584 CA Utrecht, The Netherlands}
\affiliation{Kavli Institute of NanoScience, Faculty of Applied Sciences, Delft University of Technology, Lorentzweg 1, 2628 CJ Delft, The Netherlands}

\author{J.J.A. Baselmans}
\email{j.baselmans@sron.nl}

\affiliation{SRON National Institute for Space Research, Sorbonnelaan 2, 3584 CA Utrecht, The Netherlands}

\author{J. Bueno}

\affiliation{SRON National Institute for Space Research, Sorbonnelaan 2, 3584 CA Utrecht, The Netherlands}

\author{N. Llombart}

\affiliation{Terahertz Sensing Group, Faculty of Electrical Engineering, Mathematics and Computer Science, Delft University of Technology, Mekelweg 4, 2628 CD Delft, The Netherlands}

\author{T.M. Klapwijk}
\affiliation{Kavli Institute of NanoScience, Faculty of Applied Sciences, Delft University of Technology, Lorentzweg 1, 2628 CJ Delft, The Netherlands}

\date{\today}

\begin{abstract}

We report on fluctuations in the electron system, Cooper pairs and quasiparticles, of a superconducting aluminium film. The superconductor is exposed to pair-breaking photons (1.54 THz), which are coupled through an antenna. The change in the complex conductivity of the superconductor upon a change in the quasiparticle number is read out by a microwave resonator. A large range in radiation power can be chosen by carefully filtering the radiation from a blackbody source. We identify two regimes. At high radiation power, fluctuations in the electron system caused by the random arrival rate of the photons are resolved, giving a straightforward measure of the optical efficiency ($48\pm 8$\%). At low radiation power fluctuations are dominated by excess quasiparticles, the number of which is measured through their recombination lifetime. 

\end{abstract}

\maketitle

In a superconductor well below its critical temperature, the majority of the electrons is bound in a condensate of Cooper pairs. The further the superconductor is cooled down, the closer it gets to its ground state, where all the quasiparticles are condensed to pairs. Due to the low gap energy, the superconductor is sensitive to disturbances from the environment to which it couples. In most experiments this sensitivity is undesirable, but it is particularly suited for detection of radiation. The superconductor can interact with its environment due to either photons or phonons. Photons with an energy higher than the energy gap break up Cooper pairs into quasiparticles. The change in the number of quasiparticles and Cooper pairs changes the electrodynamic response of the superconductor, which can be measured using a microwave resonator \cite{pday2003}. Quasiparticles give rise to microwave losses and the Cooper pairs to a kinetic inductance \cite{dmattis1958}. In steady state, the number of quasiparticles fluctuates in time around a constant average value. A measurement of the spectrum of these fluctuations allows for a characterisation of the quasiparticle system when exposed to pair-breaking photons, microwave photons or to changes in the bath temperature. The characteristic timescale of the fluctuations, the quasiparticle recombination time, is inversely proportional to the number of quasiparticles \cite{skaplan1976}, and is therefore a measure of this number. These fluctuation phenomena are a monitor of the superconducting state and reveal the physical mechanisms that are at the heart of pair breaking in a superconductor.

\begin{figure*}

\includegraphics[width=0.9\textwidth]{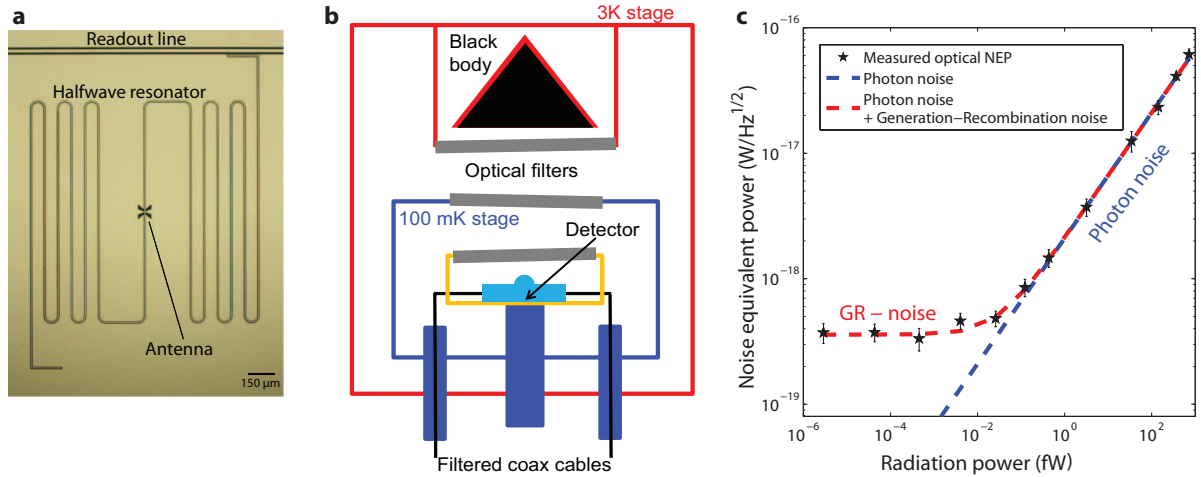}
\caption{\label{fig:detector} \textbf{Overview of the experiment and main results.} \textbf{a}, A picture of one antenna-coupled microwave resonator. It consists of a half wavelength coplanar waveguide microwave resonator with two open ends, capacitively coupled to a microwave readout line. In the middle it has an X-slot antenna to receive optical radiation. The ground plane and central strip of the resonator are respectively 100 and 50 nm thick layers of aluminium. \textbf{b}, Schematic of the setup for measurements at various radiation powers. A blackbody with a variable temperature illuminates the lens-antenna coupled resonators through three stacks of filters, which define a passband around 1.54 THz. Because of the box-in-box configuration at 100 mK and the coax cable filters, the device is well shielded from stray light. \textbf{c}, The optical sensitivity, expressed in Noise Equivalent Power (NEP) as a function of radiation power at a frequency of 20 Hz. We observe two regimes: above 0.1 fW the NEP increases with $\sqrt{P_{rad}}$, indicative of photon noise, whereas below 0.1 fW the NEP saturates. The blue line is the photon noise limit as a function of power, with the optical efficiency (48\%) taken into account. For the red dashed line, the generation-recombination noise limit due to excess quasiparticles is taken into account, based on a quasiparticle recombination time of 3 ms. The error bars are combined statistical uncertainties from the noise level and responsivity. }

\end{figure*}

We study these processes in a superconducting pair breaking detector formed by a 50 nm thick Al film. The ideal pair breaking detector can either count single photons while its sensitivity is limited by Fano noise, or is photon integrating and limited by photon noise, the noise from the photon source itself \cite{syates2011}. In both cases a high optical efficiency is required. The principle of radiation detection due to pair breaking with superconducting microwave resonators was proposed in 2003 \cite{pday2003,jzmuidzinas2012}. Since then, several unanticipated sources of excess noise have been identified and studied in depth \cite{jgao2007,jgao2008b,rbarends2008,rbarends2010b}. Here we report on an all-aluminium antenna-coupled microwave resonator detector (Fig. \ref{fig:detector}a), which is limited \emph{only} by fluctuations in the electron system that are fundamentally connected to the physical process of pair breaking. We use a blackbody with a variable temperature (3-25 K) and eight optical filters, which define an optical band around 1.54 THz, as shown schematically in Fig. \ref{fig:detector}b. The radiation power can be changed from $3\times 10^{-21}$ W to $7\times 10^{-13}$ W. At powers ranging from 0.1 - 700 fW, the sensitivity is only limited by the random arrival rate of the photons, which is evident through the measured power dependence of the noise equivalent power (NEP) as shown in Fig. \ref{fig:detector}c. At lower radiation powers, we observe a power-independent NEP. This is consistent with generation-recombination noise due to the presence of excess quasiparticles \cite{cwilson2001,pdevisser2011}. Excess quasiparticles are a general concern for superconducting devices \cite{jmartinis2009,gcatelani2011,mlenander2011,mzgirski2011,rbarends2011,osaira2012,driste2012}. In this case they are generated by the microwave readout power \cite{pdevisser2012b,dgoldie2013}. As shown in Fig. \ref{fig:detector}c, they impose a lower limit to the NEP of this detector of $3.8\pm 0.6\times 10^{-19}$ WHz$^{-1/2}$, which is the lowest reported so far for this type of detectors.\\

\textbf{Design of the experiment.}
The detector is based on a lens-antenna coupled superconducting microwave resonator. The resonator is an open ended half wave, coplanar waveguide resonator, where the central strip (with a width of 3 $\mu$m) is isolated from the ground plane. The resonators all have different lengths and therefore different resonant frequencies, enabling the read-out of all resonators using a single coaxial line. Radiation is focused by a silicon lens to an X-slot antenna \cite{aiacono2011}, optimised for broad band detection from 1.4-2.8 THz. Radiation coupled to the antenna is launched as a travelling wave into the waveguide \cite{aiacono2011}, where it is absorbed by breaking Cooper pairs (the gap energy $\Delta=$ 188 $\mu$eV). The created quasiparticles, which can diffuse over several millimeters before they recombine, are confined to the central strip. The central strip layer is 50 nm thick, and the groundplane layer 100 nm. The thin central strip layer gives higher response and ensures that most of the radiation is absorbed in that central strip, due to its higher resistance (see Methods). The thick groundplane reduces antenna losses. An advantage of the geometry, shown in Fig. \ref{fig:detector}a, is that it can also be used at other radiation frequencies by only changing the antenna.

\begin{figure*}

\includegraphics[width=0.99\textwidth]{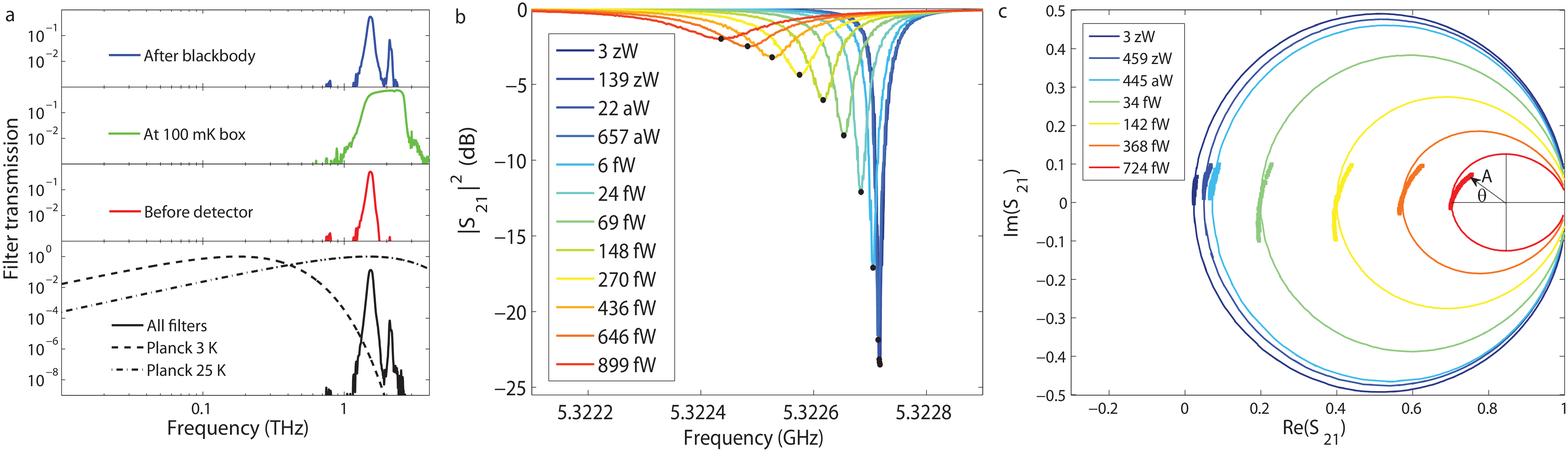}

\caption{\label{fig:filtersS21} \textbf{Response to radiation.} \textbf{a}, Filter transmission characteristics of the three stacks of optical filters in the setup (Fig. \ref{fig:detector}b). The first and third set of filters have a low-pass, a band-pass and a high-pass filter. The second set (at the 100 mK box) has only a high- and a low-pass filter. In the bottom panel the total transmission of these eight filters is shown. We also show the normalised spectral radiance (Planck's law) at two blackbody temperatures, which demonstrate the large tunability in radiation power in this spectral range. Note that especially for low blackbody temperatures only a fraction $10^{-6}$ of the total power is in the spectral range of interest. The rejection of the rest of the power requires the eight consecutive filters. \textbf{b}, The magnitude of the microwave transmission $|S_{21}|^2$, measured as a function of frequency for various radiation powers as shown in the legend. At higher power, more quasiparticles are created, which give a higher resistance and inductance and therefore lead to a lower resonant frequency and a shallower dip. The dots show the resonant frequency at each power. \textbf{c}, The resonance circle for a selection of radiation powers (legend), measured as a function of frequency (lines). The squares show the response upon a small change in the radiation power measured at constant frequency, the resonant frequency of each circle. In the last circle we show how that response is translated into an amplitude, $A$, and a phase, $\theta$.}

\end{figure*}

The sample is cooled in a pulse tube precooled adiabatic demagnetization refrigerator. The sample stage is carefully shielded from stray light from the 3 K stage of the cooler, using a box-in-a-box concept with optical filters at each stage, as well as coax cable filters in the outer box \cite{jbaselmans2012}. The photon source is a blackbody with a variable temperature between 3 and 25 K. The system is schematically depicted in Fig. \ref{fig:detector}b. Eight optical filters in series define an optical bandpass of 0.1 THz centred around 1.54 THz. Three filter stacks are essential to eliminate filter heating. The filter transmission of the three filter stages is shown in Fig. \ref{fig:filtersS21}a. The curves of spectral radiance for high and low blackbody temperature indicate a large tuning range in radiation power ($P_{rad}$). In fact, $P_{rad}$ can be varied between 3 zW (1 zW = $10^{-21}$ W) and 1 pW (Supplementary Information). Practically this experiment allows us to switch from a regime where the number of quasiparticles is fully determined by the radiation to a regime with a negligible number of optically created quasiparticles. We put a polarising wire grid just before the detector to make sure the detector only receives radiation in the polarisation direction of the antenna.\\

\textbf{Operation principle.}
The number of quasiparticles is measured through a measurement of the complex conductivity of the superconductor. The real part of the conductivity, $\sigma_1$, is due to the quasiparticles and resistive. The imaginary part, $\sigma_2$, is due to the kinetic inductance of the Cooper-pair condensate \cite{dmattis1958}. When the radiation power or the bath temperature is increased, more quasiparticles are generated, which changes both $\sigma_1$ and $\sigma_2$. The kinetic inductance increases, which leads to a lower resonant frequency $f_0=1/4l\sqrt{(L_g+L_k)C}$, where $l$ is the length of the resonator, $L_g$ the geometrical inductance, $L_k$ the kinetic inductance and $C$ the capacitance of the line, all per unit length. The losses at microwave frequencies also increase, leading to a shallower resonance. Measurements of the resonance curves for various radiation powers are shown in Fig. \ref{fig:filtersS21}b. In a practical detection scheme one typically uses an amplitude, $A$, and a phase, $\theta$, referred to the resonance circle in the complex plane \cite{jgao2007}, as shown in Fig. \ref{fig:filtersS21}c. The amplitude response originates from a change in resistance, whereas the phase changes due to the kinetic inductance. We have only used the amplitude response in this experiment.

The NEP is a convenient quantity to compare the spectra of quasiparticle fluctuations in different regimes, as shown in Fig. \ref{fig:detector}c. The NEP of the resonator amplitude is experimentally determined from a measurement of the noise spectrum ($S_A$) and the responsivity to radiation ($dA/dP_{rad}$) and given by
\begin{equation}
	NEP(f)=\sqrt{S_A(f)}\left(\frac{dA}{dP_{rad}}\right)^{-1}\sqrt{1+(2\pi f\tau_{qp})^2},
\label{eq:measuredNEP}
\end{equation}
with $P_{rad}$ the radiation power and $f$ the modulation frequency. $dA/dP_{rad}$ is obtained experimentally by a linear fit to a measurement of $A$ where $P_{rad}$ was slowly varied around the power of interest (Fig. \ref{fig:filtersS21}c). The last factor in Eq. \ref{eq:measuredNEP} arises because the quasiparticle system cannot respond to fluctuations that are faster than the quasiparticle recombination time, $\tau_{qp}$. \\

\begin{figure*}

\includegraphics[width=.63\textwidth]{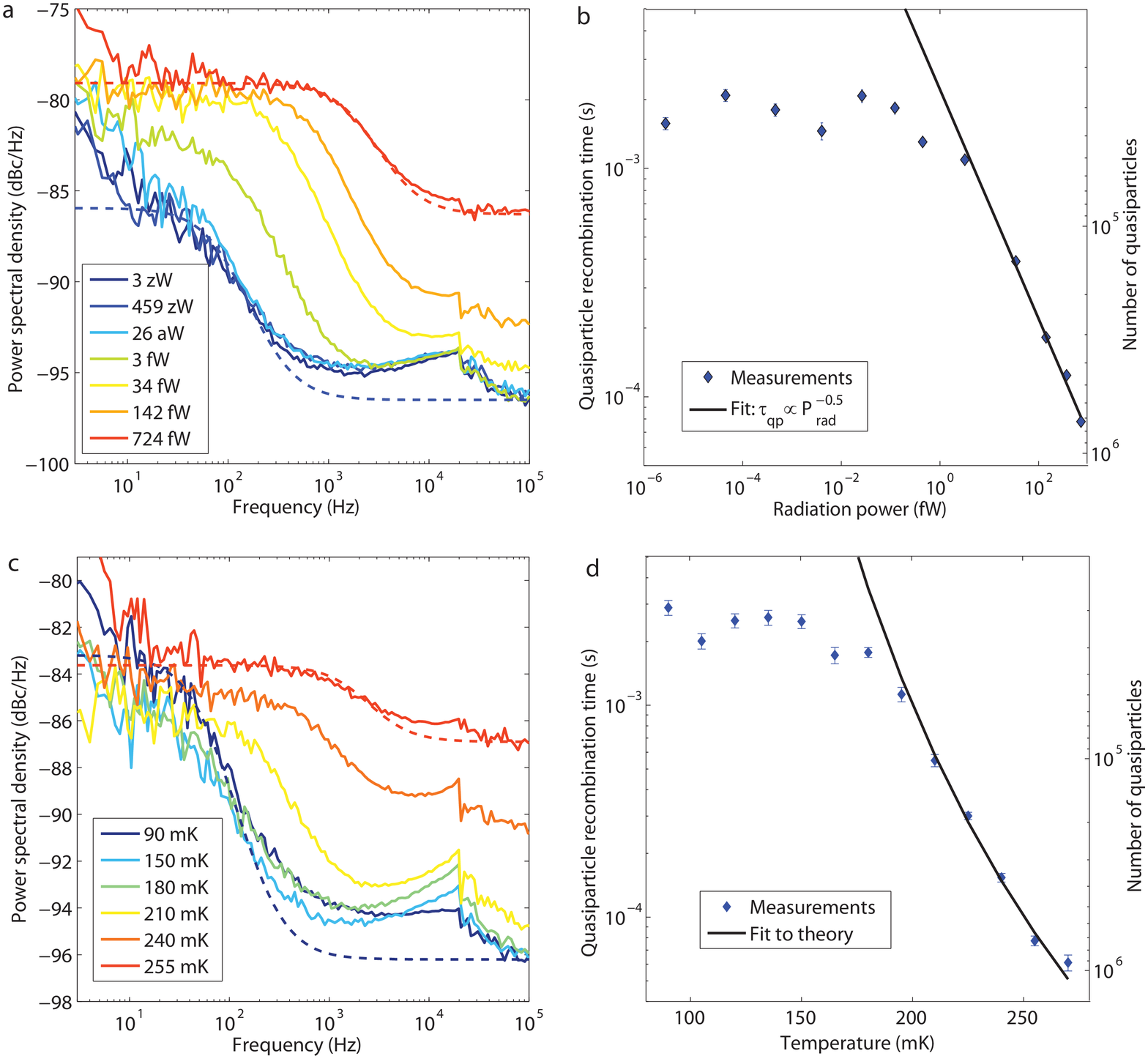}

\caption{\label{fig:fluctuations} \textbf{Quasiparticle fluctuations.} \textbf{a}, Power spectral density of the resonator amplitude as a function of frequency for different radiation powers at a constant bath temperature of 120 mK and a microwave readout power of -88 dBm. Lorentzian fits to the spectra at the lowest and highest temperatures (dashed lines) show how the quasiparticle recombination time can be extracted from the spectra. A noise floor due to amplifier noise is added to the fitted roll-off. \textbf{b}, The quasiparticle recombination time as a function of radiation power obtained from the roll-off frequency in the measured spectra. The error bars denote statistical uncertainties from the fitting procedure. The fit is a power law to the last five points (where $\tau_{qp}$ does not saturate): $\tau_{qp} \propto P_{rad}^{-0.50\pm 0.03}$. The right axis shows the number of quasiparticles corresponding to the measured recombination time. \textbf{c}, Power spectral density of the resonator amplitude as a function of frequency for different bath temperatures at a microwave readout power of -88 dBm. As expected the level of the spectrum stays constant and the roll-off frequency increases with increasing temperature, corresponding to a decreasing recombination time. At the highest two temperatures, the spectral level starts to rise, because the amplifier noise starts to dominate. \textbf{d}, Quasiparticle recombination time as a function of temperature as extracted from the spectra. The solid line is the theoretical expectation for the recombination time from Ref. \onlinecite{skaplan1976}. The right axis shows the number of quasiparticles corresponding to the measured recombination time. }

\end{figure*}

\textbf{Photon-induced quasiparticle fluctuations.}
If the average number of quasiparticles is dominated by the absorbed optical photons, the number of quasiparticles fluctuates in time due to two contributions. One is fundamental to every power-integrating detector and due to the random arrival rate of the photons, which induce a random generation of quasiparticles. The power spectral density of fluctuations in the resonator amplitude due to this photon noise is given by \cite{syates2011}
\begin{equation}
	S_{A}^P(f) = 2hFP_{rad}(1+mB)\frac{\left(dA/dP_{rad}\right)^2}{1+(2\pi f\tau_{qp})^2},
\label{eq:radiusphotonnoise}
\end{equation}
where the first term is the spectrum of the photon (power) fluctuations and the second term describes the resonator response upon a change in the radiation power. $F$ is the frequency of the optical photons and $h$ Planck's constant. The factor $(1+mB)$ is the correction to Poissonian statistics due to photon bunching, with $m$ the efficiency from emission to detection of one mode and $B$ the mode occupation \cite{rboyd1982}, which is negligible for the here measured power range. Eq. \ref{eq:radiusphotonnoise} is valid as long as $\tau_{qp}\gg\tau_{res}$, which holds in this experiment since the response time of the resonator, given by $\tau_{res}=Q/\pi f_0$, is 6 $\mu$s. $Q$ is the quality factor of the resonator. 

Because of the pair-breaking nature of the radiation absorption a second noise mechanism arises due to random recombination of the quasiparticles that are generated by the photons. This is half the generation-recombination noise that arises in thermal equilibrium \cite{cwilson2004}, because generation noise is already contained in Eq. \ref{eq:radiusphotonnoise}. The spectrum is given by
\begin{equation}
	S_{A}^R(f) = \frac{2N_{qp}\tau_{qp}}{1+(2\pi f\tau_{qp})^2}\left(\frac{dA}{dN_{qp}}\right)^2,
\label{eq:radiusrecnoise}
\end{equation}
with $N_{qp}$ the number of quasiparticles and $dA/dN_{qp}$ the responsivity of $A$ to a change in $N_{qp}$. Quasiparticle number fluctuations can be converted to power fluctuations through $\eta_{pb}\eta_{opt}P_{rad}=N_{qp}\Delta/\tau_{qp}$. $\eta_{opt}$ is the optical efficiency, the efficiency with which power in front of the lens is absorbed in the detector. $\eta_{pb}\approx 0.6$ is the pair breaking efficiency \cite{akozorezov2000}, the efficiency with which absorbed radiation power is converted into quasiparticles. For small changes in the quasiparticle number, $dP_{rad}/dN_{qp}=\Delta/\tau_{qp}\eta_{pb}\eta_{opt}$ and therefore $dA/dP_{rad} = \tau_{qp}\eta_{pb}\eta_{opt}/\Delta\cdot(dA/dN_{qp})$. From Eqs. \ref{eq:radiusphotonnoise} and \ref{eq:radiusrecnoise}, the relative contribution of photon noise compared to recombination noise is given by $hF(1+mB)\eta_{pb}\eta_{opt}/\Delta=10$ at all $P_{rad}$, for $F=1.54$ THz and $\eta_{opt} = 0.5$. 

The NEP due to photon noise and recombination noise (Eqs. \ref{eq:measuredNEP}-\ref{eq:radiusrecnoise}), for $f<1/(2\pi\tau_{qp})$, is given by 
\begin{equation}
	NEP_{photon} = \sqrt{\frac{2P_{rad}hF(1+mB)+2\Delta P_{rad}/\eta_{pb}}{\eta_{opt}}},
\label{eq:opticalNEP}
\end{equation}
which is shown as the blue dashed line in Fig. \ref{fig:detector}c.

In thermal equilibrium $N_{qp}$ is related to $\tau_{qp}$ through
\begin{equation}
	N_{qp} = \frac{1}{\tau_{qp}}\frac{\tau_0N_0(k_BT_c)^3V}{2\Delta^2},
\label{eq:Nqptauqp}
\end{equation}
where $N_0$ is the single spin density of states at the Fermi level, $V$ the volume and $\tau_0$ the characteristic electron-phonon interaction time \cite{skaplan1976}. We take $N_0=1.72\times 10^{10}$ $\mu$m$^{-3}$ eV$^{-1}$ and $V=0.6\times 10^3$ $\mu$m$^{3}$, half the volume of the central strip of the resonator (see Methods). This equation is also expected to hold in non-equilibrium conditions due to optical excitations \cite{jgao2008c} or microwave readout power dissipation \cite{dgoldie2013} at low bath temperatures.

The fluctuations in the resonator amplitude (Fig. \ref{fig:filtersS21}c) were measured as a function of radiation power at a constant bath temperature of 120 mK (Methods). The resulting power spectral densities are shown in Fig. \ref{fig:fluctuations}a for a selection of radiation powers and a microwave readout power (the power on the readout line) of -88 dBm. We observe that the spectra show a roll-off, the frequency of which increases as a function of radiation power, due to the decreasing quasiparticle recombination time. The other phenomena in the noise spectrum at higher frequencies (the bump at 20 kHz, a second higher frequency roll-off and amplifier noise) are understood and can be accounted for (see Supplementary Information). The quasiparticle recombination times from the roll-off in the spectra are shown as a function of radiation power in Fig. \ref{fig:fluctuations}b. $N_{qp}$, calculated using Eq. \ref{eq:Nqptauqp} is shown on the right axis. Since $\eta_{pb}\eta_{opt}P_{rad}=N_{qp}\Delta/\tau_{qp}$ and $N_{qp}\propto 1/\tau_{qp}$ (Eq. \ref{eq:Nqptauqp}), $\tau_{qp}$ is expected to scale as $\tau_{qp} \propto P_{rad}^{-1/2}$. A fit to the measured recombination time as a function of $P_{rad}$ results in $\tau_{qp} \propto P_{rad}^{-0.50\pm 0.03}$, which agrees very well with the expected behaviour. Within the measurement accuracy the same coefficient is measured for other microwave readout powers. The quasiparticle recombination time saturates below about 0.1 fW, which is consistent with the presence of excess quasiparticles \cite{pdevisser2011}.\\

\textbf{Phonon-induced quasiparticle fluctuations.}
Excess quasiparticles give rise to quasiparticle number fluctuations \cite{pdevisser2011,pdevisser2012b}. To verify that the spectra in the saturation regime show these fluctuations, we change the number of quasiparticles by varying the number of phonons (the bath temperature) at the same microwave power. The amplitude spectrum is shown for bath temperatures ranging from 90-255 mK in Fig. \ref{fig:fluctuations}c. The blackbody temperature is kept at 3.2 K, so there are less than 100 quasiparticles due to the radiation power in the sensitive volume. The amplitude spectrum due to quasiparticle fluctuations can be described as \cite{cwilson2001,pdevisser2011} $S_{A}^{GR}(f) = 2S_{A}^{R}(f)$, because here both generation and recombination are considered. The noise level, which is proportional to $N_{qp}\tau_{qp}$, is expected to be constant as a function of temperature (see Eq. \ref{eq:Nqptauqp}), which we indeed observe in Fig. \ref{fig:fluctuations}c. We assume here that $dA/dN_{qp}$ is constant for this temperature range \cite{jgao2008c}. The quasiparticle recombination time extracted from these spectra is plotted as a function of temperature in Fig. \ref{fig:fluctuations}d. We observe the same saturation level in the recombination time as in Fig. \ref{fig:fluctuations}b where the radiation power was changed. $\tau_0=303\pm 14$ ns is obtained from a fit to the measured $\tau_{qp}$ as a function of temperature \cite{skaplan1976,pdevisser2011}. $\tau_0$ is slightly different from earlier results (458 ns \cite{pdevisser2011}), which could be due to the higher resistivity and $T_c$ of the Al \cite{skaplan1976}.

We have now verified that the noise spectra in the regime of low radiation power (below 0.1 fW) are consistent with quasiparticle number fluctuations. The optical NEP due to quasiparticle number fluctuations is given by
\begin{equation}
	NEP_{GR} = \frac{2\Delta}{\eta_{pb}\eta_{opt}}\sqrt{\frac{N_{qp}}{\tau_{qp}}}.
\label{eq:GRnoiseNEP}
\end{equation} 
The excess quasiparticles that cause the measured saturation in the recombination time (Fig. \ref{fig:fluctuations}b) are the explanation of the saturation of the NEP in Fig. \ref{fig:detector}c.

If we return to the photon induced fluctuations in Fig. \ref{fig:fluctuations}a, we observe that the noise level becomes also constant at the highest radiation powers. This constant level is expected when Eqs. \ref{eq:radiusphotonnoise} and \ref{eq:radiusrecnoise} are rewritten in terms of $N_{qp}\tau_{qp}$. The noise level is higher than in Fig. \ref{fig:fluctuations}c, because there is both photon noise and recombination noise.

\begin{figure}

\includegraphics[width=0.45\textwidth]{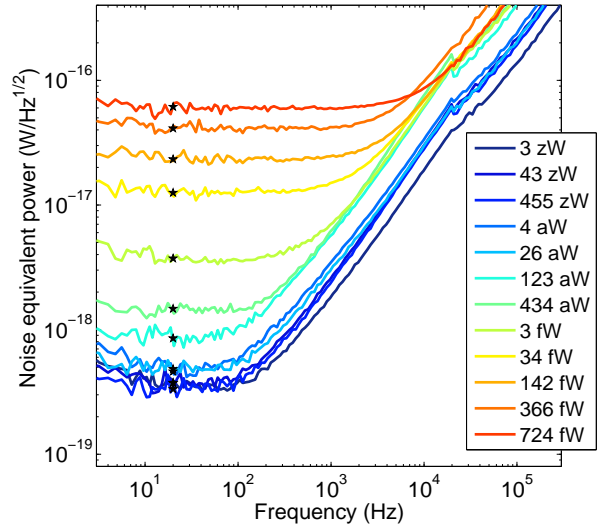}

\caption{\label{fig:NEP} \textbf{Noise equivalent power.} Measured optical noise equivalent power (NEP) in the resonator amplitude as a function of frequency for different radiation powers. The corner frequency of each spectrum corresponds to the quasiparticle recombination time as plotted in Fig. \ref{fig:fluctuations}b. The stars indicate the NEP at the reference frequency of 20 Hz, which is chosen well within the quasiparticle roll-off. These are the NEP values shown in Fig. \ref{fig:detector}c. The measurements shown are taken at the readout power that gives the minimum NEP for that radiation power.}

\end{figure}

\textbf{Noise equivalent power.}
The measured NEP, obtained by using Eq. \ref{eq:measuredNEP} together with the measured $S_A$ (Fig. \ref{fig:fluctuations}a), $dA/dP_{rad}$ (Fig. \ref{fig:filtersS21}c) and $\tau_{qp}$ (Fig. \ref{fig:fluctuations}b), is shown for various radiation powers in Fig. \ref{fig:NEP}. The NEP measurement was done at a range of microwave readout powers. The results shown in Figs. \ref{fig:detector}c and \ref{fig:NEP} are at the readout power with the minimum NEP for that radiation power. 

The measured optical NEP at 20 Hz is shown as a function of radiation power in Fig. \ref{fig:detector}c as our main result. At radiation powers of 0.1 fW and higher, the NEP scales with $\sqrt{P_{rad}}$, as expected from the photon noise limit given by Eq. \ref{eq:opticalNEP}. In this regime, the optical efficiency is obtained by fitting Eq. \ref{eq:opticalNEP} to the measured NEP. The result is shown as the blue line in Fig. \ref{fig:detector}c, which gives $\eta_{opt} = 0.48 \pm 0.08$ for a single polarisation, consistent with electromagnetic simulations of the antenna (Supplementary Information).

\begin{figure}

\includegraphics[width=0.45\textwidth]{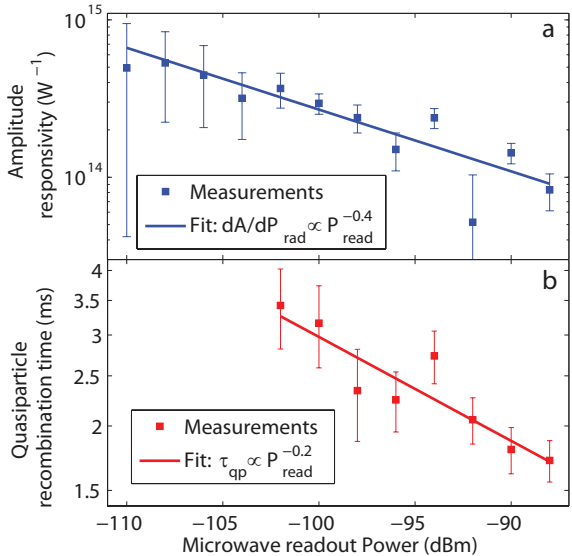}

\caption{\label{fig:Pread} \textbf{Readout power dependence.} \textbf{a}, The optical responsivity, as obtained from a fit to the measured amplitude response, as a function of microwave readout power. The line is a power law fit to the data: $dA/dP_{rad}\propto P_{read}^{-0.4\pm 0.1}$. \textbf{b}, The quasiparticle recombination time as obtained from the roll-off in the noise spectra as a function of microwave readout power. Below -102 dBm, the roll-off due to the recombination time is not visible anymore. The line is a fit to the data with a power law coefficient of $\tau_{qp}\propto P_{read}^{-0.2\pm 0.1}$. All data is measured at a radiation power of 3 zW and a bath temperature of 120 mK. }

\end{figure}

Below 0.1 fW, the NEP saturates at $3.8\pm 0.6\times 10^{-19}$ WHz$^{-1/2}$. We have seen that generation-recombination noise due to excess quasiparticles dominates the noise spectra in this regime. From the measured recombination time (3 ms, see Fig. \ref{fig:Pread}b), we calculate $N_{qp}$ using Eq. \ref{eq:Nqptauqp}. The sum of Eqs. \ref{eq:opticalNEP} and \ref{eq:GRnoiseNEP} is shown as the red dashed line in Fig. \ref{fig:detector}c and gives a good account of the measured NEP. The limit of $3.8\times 10^{-19}$ WHz$^{-1/2}$ is in good agreement with predictions based on dark experiments \cite{pdevisser2012b}. \\

\textbf{Excess quasiparticles due to the readout power.}
From recent dark experiments \cite{pdevisser2012b} (without blackbody source) and simulations \cite{dgoldie2013} we expect that the excess quasiparticles at low radiation power are due to microwave readout power dissipation. In an optical experiment excess quasiparticles should lead to a decrease in the optical response. More precisely, the optical responsivity is expected to scale with readout power as $dA/P_{rad}\propto P_{read}^{-0.5}$ in the regime where the readout power dominates the number of quasiparticles (see Supplementary Information for a derivation). The measured responsivity is shown in Fig. \ref{fig:Pread}a as a function of $P_{read}$. A power law to the responsivity versus readout power results in $dA/P_{rad}\propto P_{read}^{-0.4\pm 0.1}$, which is in reasonable agreement with the expected scaling. 

Fig. \ref{fig:Pread}b shows $\tau_{qp}$, extracted from the noise spectra, as a function of $P_{read}$. $\tau_{qp}$ increases when $P_{read}$ decreases, which is consistent with quasiparticle generation by the microwave readout signal \cite{pdevisser2012b,dgoldie2013}. A fit to the measured data gives $\tau_{qp}\propto P_{read}^{-0.2\pm 0.05}$. If the absorbed microwave power would scale linearly with $P_{read}$, we would expect $\tau_{qp}\propto P_{read}^{-0.5}$ (Supplementary Information). The difference may be caused by the intricate distribution of the quasiparticle over energies due to the microwave absorption \cite{dgoldie2013}.

\textbf{Discussion.}
At higher radiation powers, where the noise spectrum is dominated by photon noise, the optical responsivity also changes with readout power. In this regime ($P_{rad}>$ 1 fW) however, the measured photon noise NEP stays the same, as expected from Eq. \ref{eq:opticalNEP} (Supplementary Information). Therefore, when photon-noise dominates the noise spectrum, one can safely use high readout powers to suppress amplifier noise.

At the lowest readout power where $\tau_{qp}$ was determined, -102 dBm, the quasiparticle recombination time is 3.5 ms, which corresponds to a quasiparticle density $n_{qp} = 24$ $\mu$m$^{-3}$. This density is still high in comparison with the lowest reported values for superconducting qubits and Cooper pair transistors \cite{osaira2012,driste2012} (less than 0.1 $\mu$m$^{-3}$), but inherent to the relatively high microwave powers we need in this type of experiments. The measured limit in optical NEP due to excess quasiparticles is comparable to the lowest observed optical NEP in other detectors for similar wavelengths \cite{bkarasik2011,maudley2012,kstone2012}. 

A reduction in $N_{qp}$ is possible by using a parametric amplifier with high bandwidth and dynamic range \cite{beom2012}. This allows a reduction of the readout power by about a factor 10. In the current design however, the detector would become too slow for practical use at low readout power due to the long recombination time. The most feasible route towards lower NEP with aluminium, the most reliable material so far, is to choose geometries in which the active volume is dramatically reduced, which could also be the route towards single photon counting at terahertz frequencies.\\

\textbf{Methods}\\
\textbf{Sample design.}
A layer of aluminium with a thickness of 100 nm is sputtered onto a sapphire substrate and serves as the ground plane for the microwave resonators. The microwave resonator is a coplanar waveguide resonator with a central strip width of 3 $\mu$m and slit widths of 1.5 $\mu$m. The central strip of the resonator is made of a second layer of 50 nm Al. The critical temperature of the 50 nm layer is measured to be $T_c = 1.24$ K, from which the energy gap $\Delta = 1.76 k_B T_c $ = 188 $\mu$eV, with $k_B$ Boltzmann's constant. From the normal state resistivity ($\rho = 2.2$ $\mu\Omega$cm for the central strip and 0.28 $\mu\Omega$cm for the groundplane) the skin depth for radiation at 1.54 THz is 60 nm in the central strip and 21 nm in the groundplane. The X-slot antenna would be ineffective for a layer thinner than the skin depth, therefore the groundplane layer is 100 nm thick. Since the microwave sheet resistance of the central line is 0.37 $\Omega$ and that of the ground plane 0.13 $\Omega$, about 73\% of the radiation is absorbed in the central line. 

The current distribution along the length of the resonator peaks at the antenna and decreases as $\sin(x)$ to zero at the open ends. Therefore the responsivity changes with $\sin^2(x)$. Since the diffusion length within a typical quasiparticle recombination lifetime of 2 ms is more than half the resonator length, optically created quasiparticles can move into the non-responsive regime. Therefore for calculating the number of quasiparticles in the sensitive volume, we take half the central strip volume, $V=0.6\times 10^3$ $\mu$m$^{3}$.

\textbf{Noise measurement.} The signal from the microwave generator is first attenuated, sent through the sample, and amplified with a HEMT amplifier at 4 K and with a room temperature amplifier. The output is mixed with the original signal using an IQ mixer, the output of which can be sampled at a maximum frequency of 2 MHz. The spectrum of fluctuations in the resonator amplitude is measured by recording the resonator amplitude as a function of time and computing the power spectral density. Peaks in the time domain stream that occur due to high energy impacts are filtered out before the spectrum is computed, as described in Ref. \onlinecite{pdevisser2011}. We use the amplitude direction because fluctuations in the phase direction are dominated by two level system noise in the dielectrics surrounding the resonator \cite{jgao2007} (Supplementary Information). \\

\textbf{Acknowledgements}\\
We would like to thank Y.J.Y. Lankwarden for fabricating the devices and S.J.C. Yates for contributing to the experimental setup. This work has been supported as part of a collaborative project, SPACEKIDS, funded via grant 313320 provided by the European Commission under Theme SPA.2012.2.2-01 of Framework Programme 7.\\


\clearpage
\onecolumngrid
\setcounter{figure}{0}

\renewcommand{\thefigure}{S\arabic{figure}}

\section*{Supplementary information for: "Fluctuations in the electron system of a superconductor exposed to a photon flux"} 

\subsection*{P.J. de Visser, J.J.A. Baselmans, J. Bueno, N. Llombart and T.M. Klapwijk}

\author{P.J. de Visser}
\email{p.j.devisser@tudelft.nl}

\affiliation{SRON National Institute for Space Research, Sorbonnelaan 2, 3584 CA Utrecht, The Netherlands}
\affiliation{Kavli Institute of NanoScience, Faculty of Applied Sciences, Delft University of Technology, Lorentzweg 1, 2628 CJ Delft, The Netherlands}

\author{J.J.A. Baselmans}

\affiliation{SRON National Institute for Space Research, Sorbonnelaan 2, 3584 CA Utrecht, The Netherlands}

\author{J. Bueno}

\affiliation{SRON National Institute for Space Research, Sorbonnelaan 2, 3584 CA Utrecht, The Netherlands}

\author{N. Llombart}

\affiliation{Terahertz Sensing Group, Faculty of Electrical Engineering, Mathematics and Computer Science, Delft University of Technology, Mekelweg 4, 2628 CD Delft, The Netherlands}

\author{T.M. Klapwijk}
\affiliation{Kavli Institute of NanoScience, Faculty of Applied Sciences, Delft University of Technology, Lorentzweg 1, 2628 CJ Delft, The Netherlands}

\date{\today}

\maketitle

In this supplementary document, we provide details on the experimental setup, derive relations that are used in the main article and provide additional experimental data.

\section{Optical system and radiation power}
\begin{figure}[h]
\includegraphics[width=0.32\columnwidth]{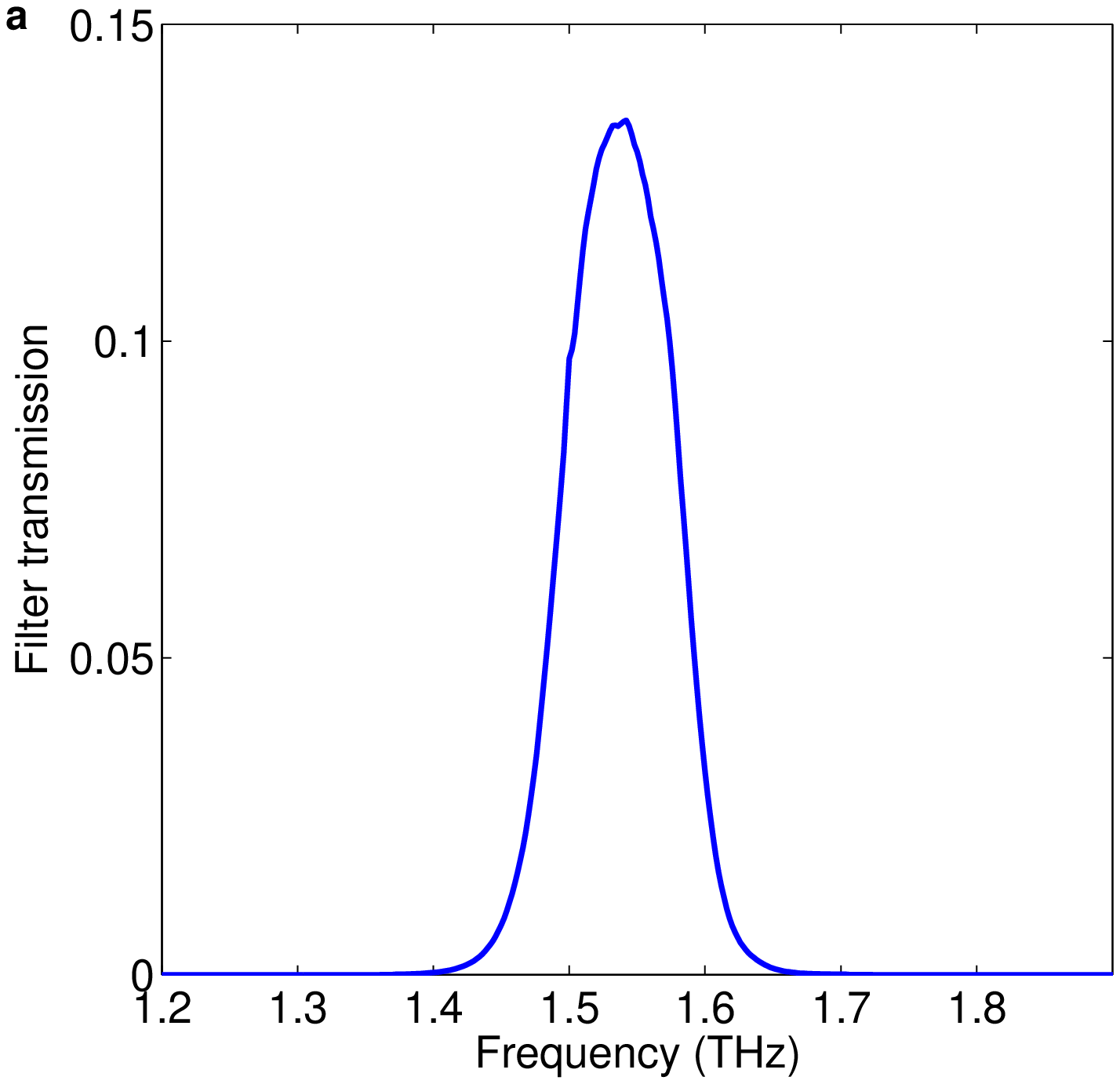}
\includegraphics[width=0.32\columnwidth]{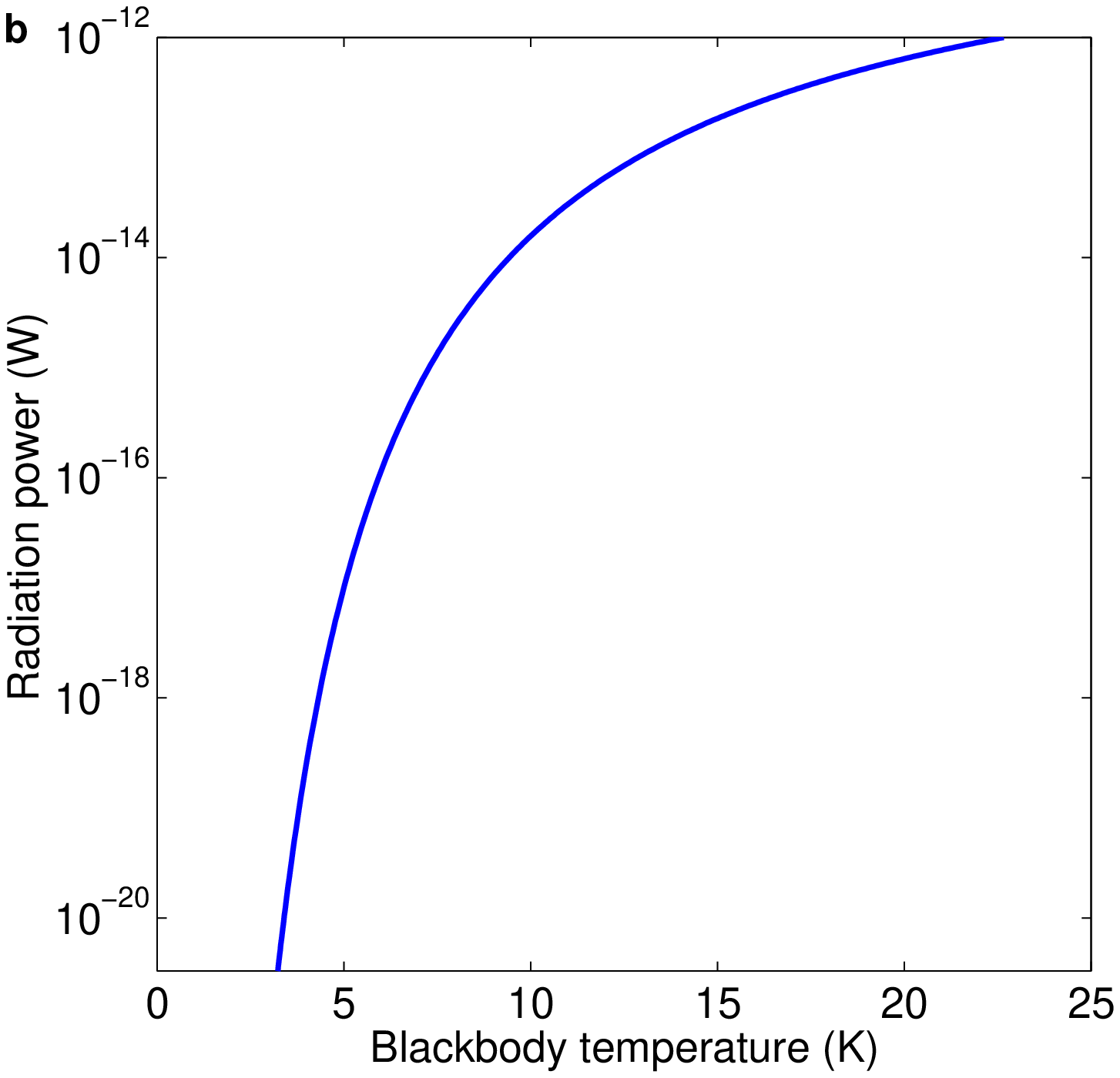}
\includegraphics[width=0.32\columnwidth]{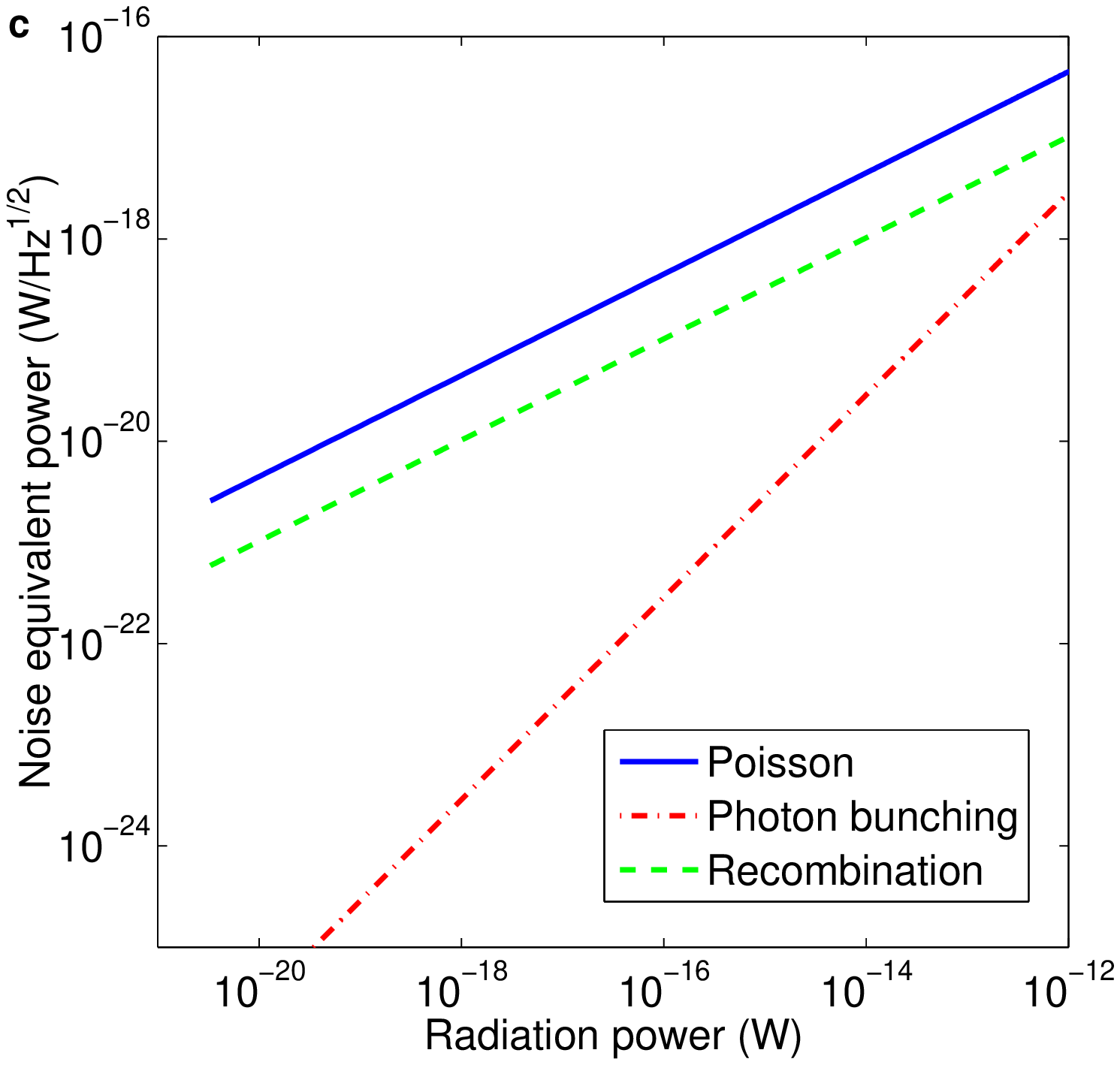}
\caption{\label{figS:filterchar} \textbf{a}, Total transmission of the filterstack as a function of frequency. \textbf{b}, Radiation power in the filter transmission band as a function of blackbody temperature. \textbf{c}, The noise equivalent power as a function of radiation power. The three lines are the contributions to the optical NEP due to the Poisson statistics of the photon stream, due to the photon bunching statistics, and due to the random recombination of quasiparticles.} 

\end{figure}

The photon source used in the experiment is a blackbody, which is formed by a 40 mm diameter copper cone, coated with carbon loaded epoxy (EPOTEK 920 1LB part A, with 3\% by weight carbon black and 3\% by weight EPOTEK 920 1LB part B), which is covered with 1 mm SiC grains. The temperature of the blackbody is varied in this experiment from 3.2 - 25 K. There are three metal-mesh filter stacks (QMC Instruments, Cardiff), the characteristics of which are given in the main article. The measured transmission of the whole filterstack as a function of optical frequency, $Tr(F)$, is shown in Fig. \ref{figS:filterchar}a (the same as in the bottom panel of Fig. 2a in the main text, but here on a linear scale).

Since there is no aperture limitation in between the blackbody and the detector, the optical throughput is assumed to be $(c/F)^2$, with $c$ the speed of light. The total radiation power that arrives in front of the lens of the detector can now be calculated by numerically integrating Planck's law over the throughput and the measured filter characteristic at each blackbody temperature $T_{BB}$. The radiation power is here given for one polarisation.
\begin{equation}
 P_{rad}(T_{BB}) = \int_0^{\infty} \frac{Tr(F)hF dF}{\exp(hF/k_bT_{BB})-1},
\label{eqS:opticalpower}
\end{equation}
where $h$ is Planck's constant and $k_B$ Boltzmann's constant.
The optical window around 1.54 THz, together with the blackbody temperature range of 3-25 K gives a large tuning range in radiation power, as shown in Fig. \ref{figS:filterchar}b. With the present device, we can verify the radiation power down to 100 aW using the measured quasiparticle recombination time (main article Fig. 5b). The trend of increasing optical responsivity and recombination time with decreasing microwave power, without any sign of saturation (main article Fig. 5a,b), suggests that the optical system is well characterised down to even lower radiation powers. The excess quasiparticles in the present device limit us to verify that.

The radiation power, calculated by Eq. \ref{eqS:opticalpower}, allows to calculate the different contributions to the noise equivalent power, as discussed in the main article. The photon-noise NEP is, given by
\begin{equation}
	NEP_{photon} = \sqrt{\frac{2P_{rad}hF + 2P_{rad}hFmB + 2\Delta P_{rad}/\eta_{pb}}{\eta_{opt}}},
\label{eqS:opticalNEP}
\end{equation}
where the first term is due to the Poisson statistics of the photon stream, the second term due to photon bunching (giving a correction to Poisson statistics) and the third term is the recombination noise of the quasiparticles. $m$ is the efficiency from emission to detection of one mode and $B$ is the mode occupation \cite{rboyd1982b}. The second term is much smaller than the Poisson term over the whole range of measured powers, as shown in Fig. \ref{figS:filterchar}c. The third term, the recombination noise, is also shown in Fig. \ref{figS:filterchar}c, which shows that the contribution due to recombination noise is small compared to photon noise, as discussed in the main article. The lines in Fig. \ref{figS:filterchar}c are calculated with $\eta_{opt}=100$\%.

Right in front of the detector, after the last optical filter, we place a polariser to select the polarization for which the antenna is designed. The polariser consists of a copper wire grid on top of a 1.5 $\mu$m thick Mylar film. The grid lines are 10 $\mu $m wide and the spacing between the lines is 20 $\mu$m.

\begin{figure}
\includegraphics[width=0.4\columnwidth]{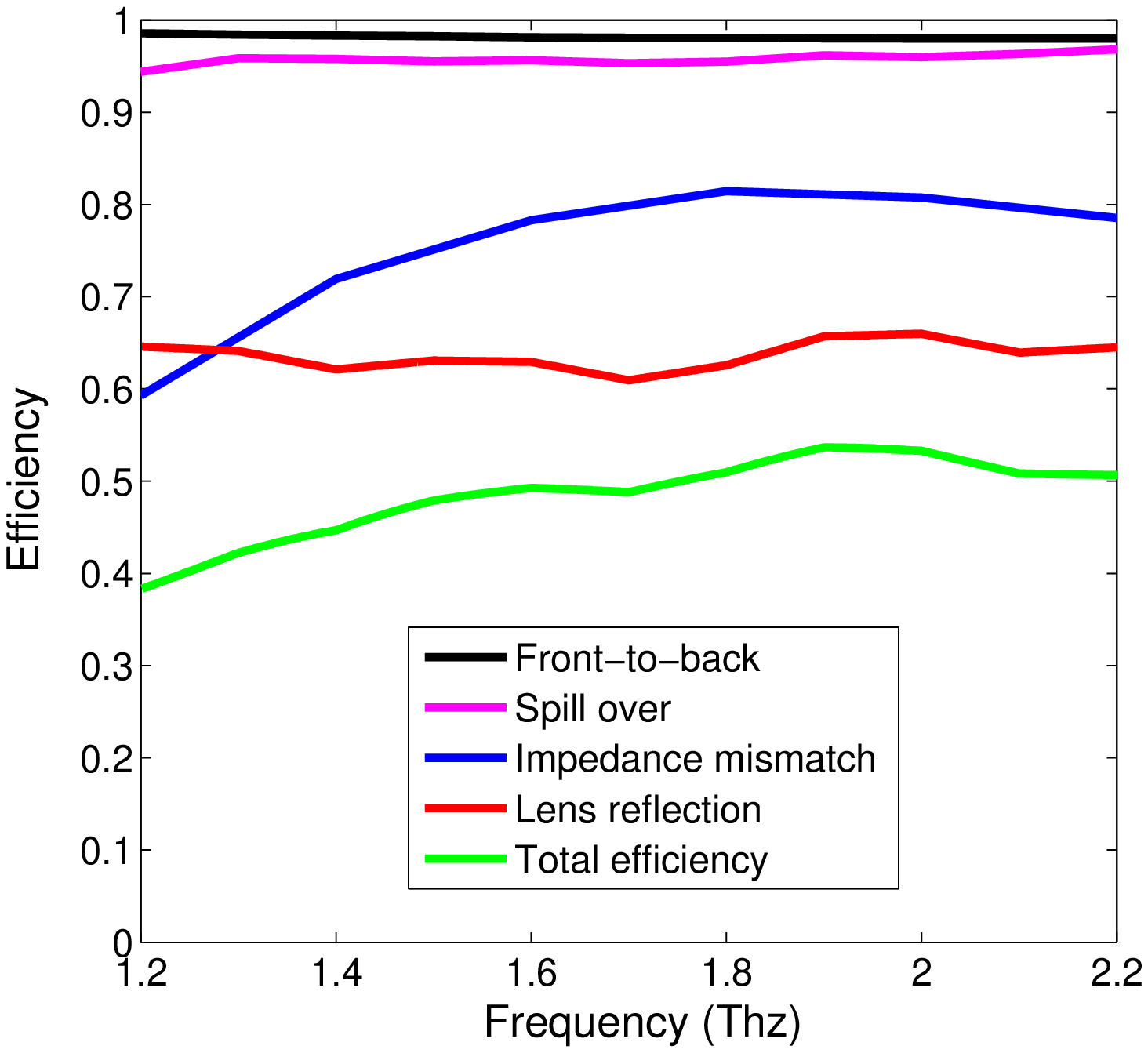}
\caption{\label{figS:CSTsim} The optical efficiency of the antenna-lens system as a function of frequency as calculated with CST Microwave Studio. The total efficiency is the multiplication of the front-to-back ratio, the spill over efficiency, the efficiency of an impedance mismatch between the antenna and the CPW line and the reflection losses at the anti-reflection coated lens surface.} 

\end{figure}

The radiation power is focused by an elliptical silicon lens of 2 mm in diameter onto the antenna, which is in the second focus of the lens \cite{dfilipovic1993b}. The major and minor axis of the ellips are 1.037 mm and 0.992 mm respectively. The lens has an anti-reflection coating of 130 $\mu$m of Parylene C, which is not optimised for 1.54 THz. The antenna is an in-line X-slot antenna, designed to receive radiation in a broad band around 1.54 THz as described in Ref. \onlinecite{aiacono2011b}. To obtain the optical efficiency, a simulation in CST Microwave Studio is performed of the whole structure: the lens with the coating, the antenna and a piece of coplanar waveguide transmission line. The aperture is chosen to be 30 degrees, the angle from which the detector can see the blackbody (single side angle). The optical efficiency is shown as a function of frequency in Fig. \ref{figS:CSTsim}. The total efficiency is the multiplication of the front-to-back ratio, the spill over losses, the efficiency of an impedance mismatch between the antenna and the CPW line and the reflection losses at the anti-reflection coated lens surface. All together, we expect an efficiency of 48\% for one polarisation in the filter transmission band, which is in good agreement with the measured optical efficiency of 48$\pm$8\%.

We like to note here that the antenna was not designed to have a perfect optical efficiency, but to have a large bandwidth. The agreement of the measured optical efficiency with the CST-simulation shows that the optical system is understood. Improving the optical efficiency is possible by adjusting the optical components.

\section{Experimental Details}

\begin{figure}

\includegraphics[width=0.32\columnwidth]{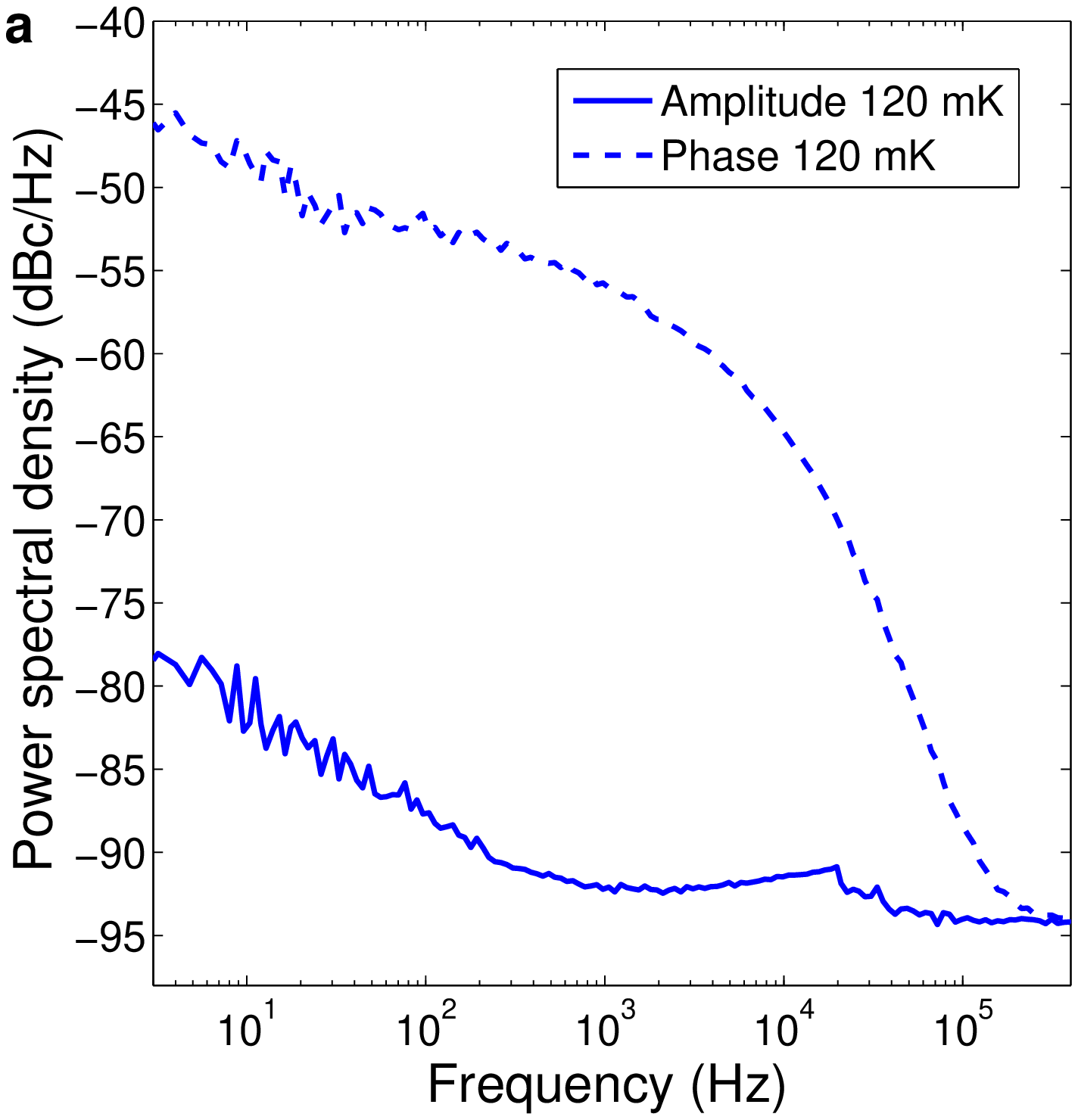}
\includegraphics[width=0.32\columnwidth]{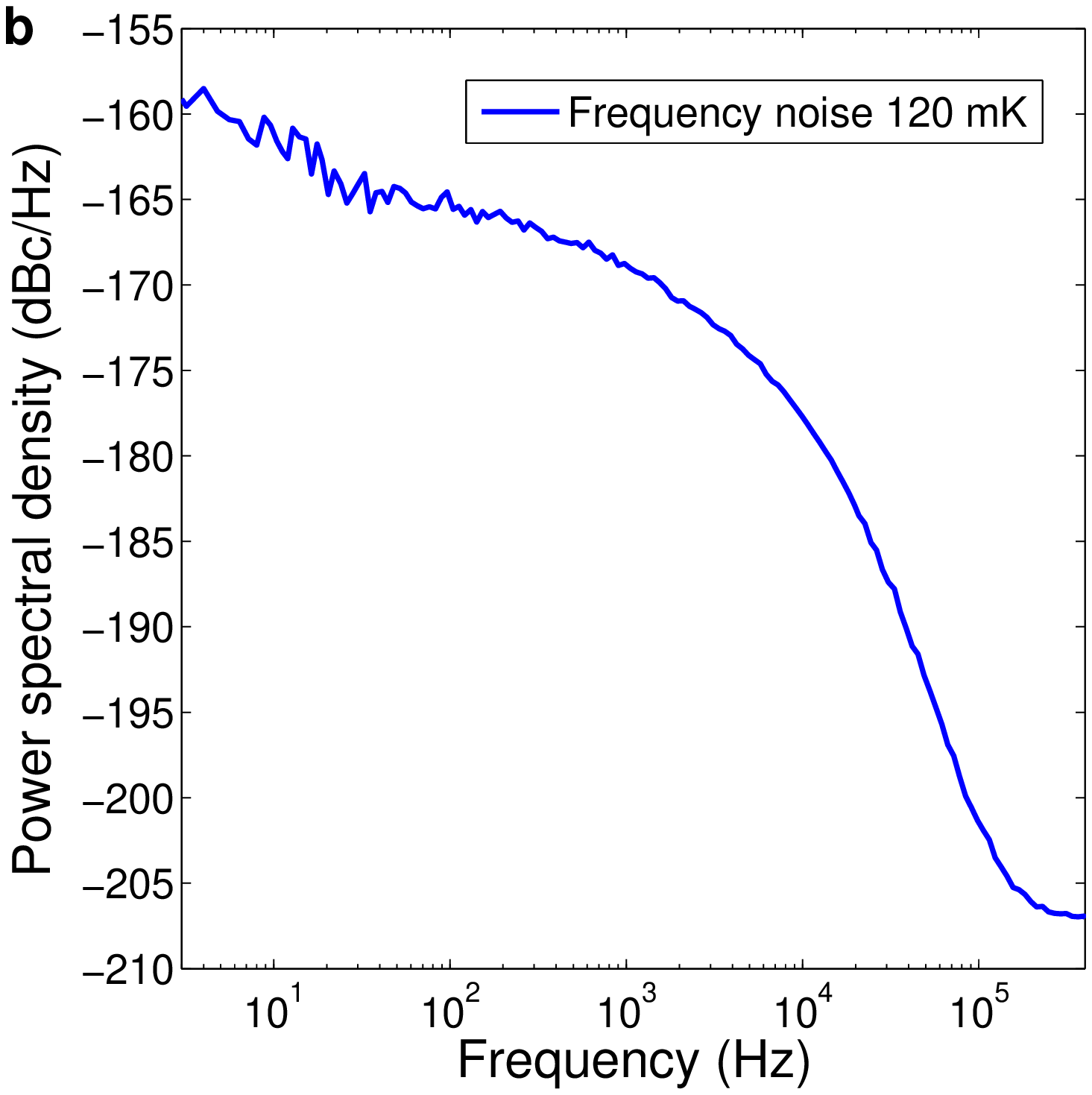}
\caption{\label{figS:phasampnoise} \textbf{a}, The power spectral density of the resonator amplitude and phase as a function of frequency at a bath temperature of 120 mK at a microwave readout power of -90 dBm. \textbf{b}, The frequency noise spectrum: $S_f = \frac{S_{\theta}}{(4Q)^2}$ at 120 mK and -90 dBm, with Q = 111711. } 

\end{figure}

In Fig. 2c of the main article we have shown that one can choose to measure the response of the superconductor in either the phase or the amplitude direction with respect to the resonance circle. Fig. \ref{figS:phasampnoise}a shows both the amplitude and phase spectra at a bath temperature of 120 mK and a microwave power of -90 dBm. Is is evident that the phase noise is 30 dB higher, due to two level system (TLS) noise \cite{jgao2007b}, which makes it impossible for this device to measure quasiparticle fluctuations in phase. Therefore we have only used the amplitude response to study quasiparticle fluctuations. To compare with previous research, we plot in Fig. \ref{figS:phasampnoise}b the frequency noise spectrum at 120 mK. The frequency noise at 1 kHz is -169 dBc/Hz, which is about 8 dB higher than reported before \cite{rbarends2009bb} for Al on sapphire, most likely due to the two layer fabrication process for this device. It is known that frequency noise decreases for higher temperatures \cite{rbarends2008b}. Therefore we have chosen to not operate at the lowest possible temperature, but at a bath temperature of 120 mK. A temperature of 120 mK is still low enough not to dominate the number of quasiparticles.

\section{Contributions to the amplitude noise spectrum}

\begin{figure}

\includegraphics[width=0.32\columnwidth]{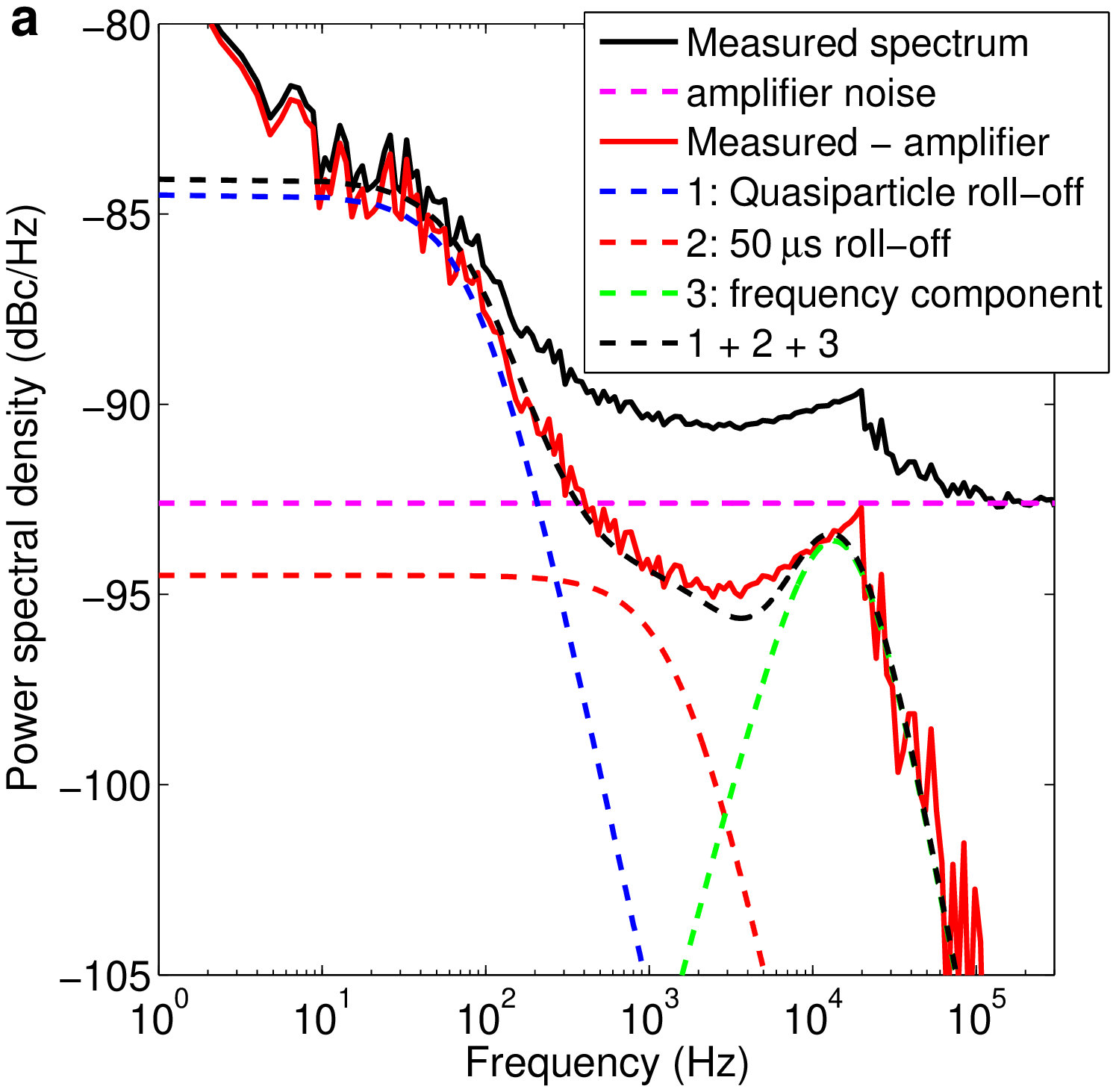}
\includegraphics[width=0.32\columnwidth]{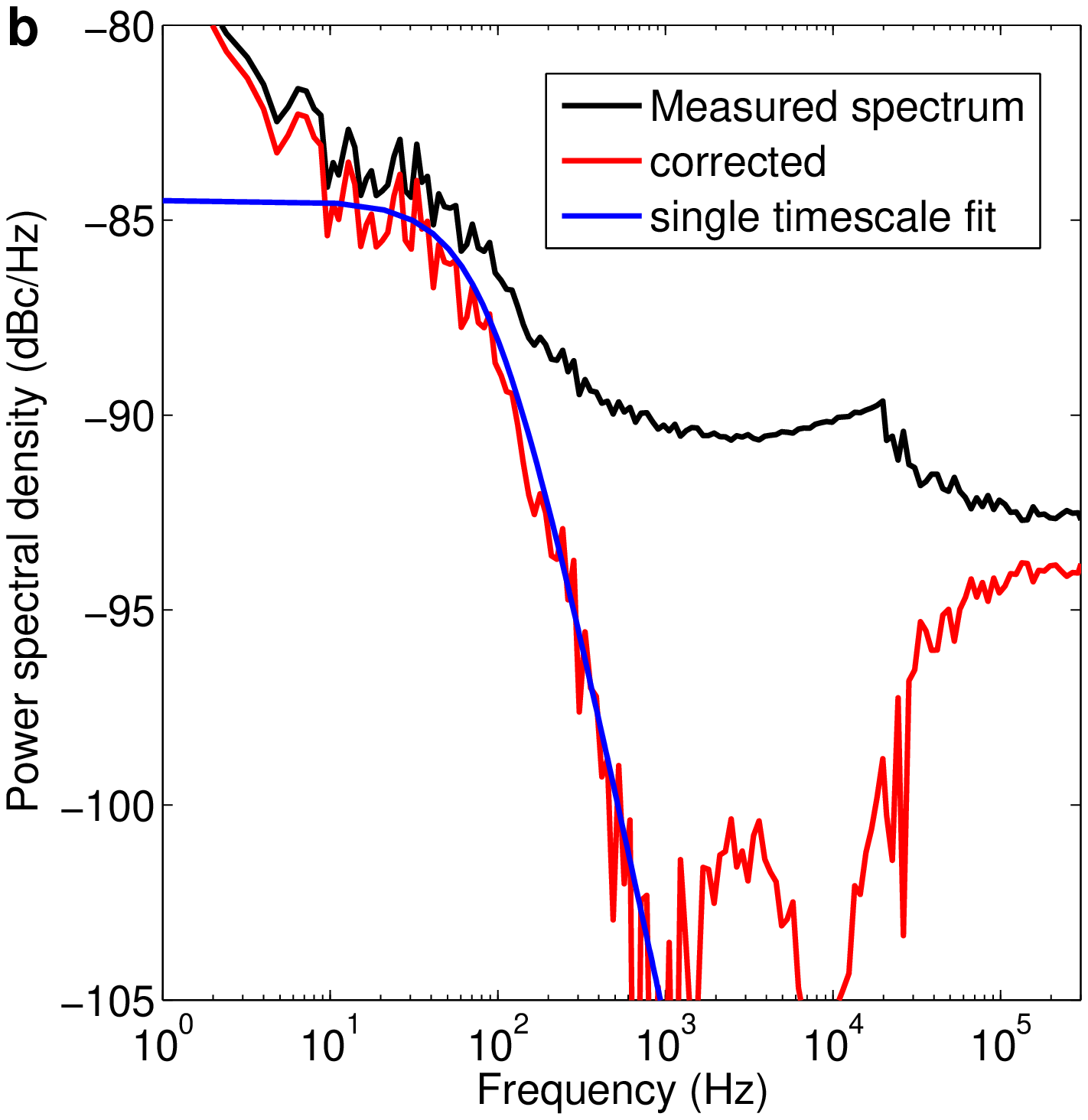}
\caption{\label{figS:noisecontributions} \textbf{a}, Different contributions to the noise spectra. The measured amplitude power spectra density at a temperature 120 mK and a microwave readout power of -92 dBm is shown as a solid black line. The amplifier noise is a white noise contribution and is determined at frequencies above 300 kHz. The measured spectrum with the amplifier noise subtracted is shown as the red line. The other dashed lines show the other contributions: the roll-off due to quasiparticle fluctuations, a second roll-off with a timescale of 50 $\mu$s and a 10 dB lower noise level, and a symmetric bump around the resonator response frequency due to mixing of frequency noise in the amplitude direction. \textbf{b}, The same measured spectrum as in \textbf{a}, together with a spectrum that is corrected by subtracting the level at 3 kHz. The correction is done to be able to only fit the quasiparticle roll-off, the result of which is shown as well. } 

\end{figure}

From Fig. 3a in the main article, it is evident that there are more contributions to the noise spectra than only the quasiparticle roll-off. In that figure, we already took into account the amplifier noise level, which will give a flat, white, noise spectrum. Fig. \ref{figS:noisecontributions}a shows as an example the measured amplitude power spectral density at the lowest radiation power and a microwave readout power of -92 dBm. Four contributions to the noise spectrum can be distinguished. Firstly, the amplifier noise, which gives a flat spectrum, the level of which can be determined through the noise level at frequencies of 200-300 kHz (-92.6 dBc/Hz in this case). The amplifier noise level is subtracted, to more clearly show the other three contributions. The second, and dominant, contribution is the roll-off due to quasiparticle fluctuations, with a level of -84.5 dBc/Hz and a characteristic time of 1.8 ms. This power spectral density has the form
\begin{equation}
	S(f) = \frac{y}{1+(2\pi f \tau)^2},
\label{eqS:rolloff}
\end{equation}
with $y$ the level and $\tau$ the timescale. The third contribution is a second roll-off of the same form, with a timescale of about 50 $\mu$s and a level which is 10 dB lower than the quasiparticle roll-off. This contribution is small and not so easily distinguishable here, but was observed more clearly in a similar resonator \cite{pdevisser2012b}. We tentatively attribute this contribution to phonon-fluctuations. The fourth contribution is a bump around the resonator response time frequency (27 kHz). It can be shown \cite{jzmuidzinasbump} that this phenomenon is consistent with mixing of frequency noise into the amplitude direction, due to a difference in the probe frequency and the resonant frequency of the resonator during the noise measurement. It can be modelled with the equation
\begin{equation}
	S(f) = y_b|\zeta(f)+\zeta^*(-f)|/4 ,
\label{eqS:bump}
\end{equation}
with $y_b$ a scaling factor and where the star denotes the complex conjugate.
\begin{equation}
	\zeta(f) = \frac{1+j\delta f_g/f_{ring}}{1+j(\delta f_g+f)/f_{ring}},
\label{eqS:bump2}
\end{equation}
with $f$ the modulation frequency, $f_{ring}=f_0/\pi Q$ the resonator ring frequency around which the bump will appear (27 kHz) and $\delta f_g$ the detuning of the generator frequency from the resonant frequency $f_0$. This detuning can occur in practice due to strong frequency noise or due to drift in either the generator frequency or the resonant frequency during the noise measurement.

Since we are only interested in modulation frequencies well within the quasiparticle recombination time bandwidth ($<$100 Hz in this case), the other noise contributions do not play a role in determining the sensitivity (NEP) of the detector. However, these contributions limit the extraction of the quasiparticle recombination time and, because they contribute mostly at higher frequencies, give a bias towards shorter lifetimes if one fits the spectra with a single-lifetime spectrum. To get a better estimate of the actual quasiparticle recombination time, we subtract from the measured noise spectrum a level which we take from noise frequencies around 1-3 kHz. The thus corrected spectrum is fitted with a single timescale Lorentzian roll-off as shown in Fig. \ref{figS:noisecontributions}b. We perform this correction because fitting all noise contributions together would require too many fit parameters. We emphasise that we only do this correction to extract a more realistic recombination time. The NEP is calculated with the measured, uncorrected, noise spectra.

\section{Derivation of the optical responsivity vs microwave power}
Here we derive how the optical responsivity of the resonator amplitude, $dA/dP_{rad}$, changes as a function of microwave readout power. We limit ourselves to the regime where the number of quasiparticles is dominated by readout power dissipation. The number of quasiparticles due to the readout power, $N_{qp}^{read}$, is related to the quasiparticle recombination time and the absorbed readout power in the quasiparticle system $P_{abs}$, through $\eta_{read}P_{abs}=N_{qp}^{read}\Delta/\tau_{qp}$. $\eta_{read}$ is the efficiency with which the absorbed microwave power creates quasiparticles. We will assume here that $P_{abs}\propto P_{read}$, with $P_{read}$ the power on the readout line (we will come back to this assumption later). Since we only derive proportionalities, we will use $P_{read}$ in the equations. $N_{qp}$ and $\tau_{qp}$ are related by \cite{skaplan1976}
\begin{equation}
	N_{qp} = \frac{\tau_0}{\tau_{qp}}\frac{N_0(k_BT_c)^3V}{2\Delta^2}=\frac{K}{\tau_{qp}},
\label{eqS:Nqptauqp}
\end{equation}
which also holds for excess quasiparticles at low temperature \cite{jgao2008c,dgoldie2013}. $K$ is one constant to replace all the other constants in this equation. In steady state $N_{qp}^{read}$ is related to $P_{read}$ as
\begin{equation}
	N_{qp}^{read}=\sqrt{\frac{\eta_{read}P_{read}}{K\Delta}}.
\label{eqS:Nqpread}
\end{equation}
We assume here that the number of quasiparticles created by the radiation $N_{qp}^{rad}$ is small (linear response regime) in which case $N_{qp}^{read}$ determines $\tau_{qp}$, ie $N_{qp}^{read}\gg N_{qp}^{rad}$. That also means $\tau_{qp}$ is expected to scale with the readout power as $\tau_{qp}\propto P_{read}^{-1/2}$. The number of quasiparticles that is created by the optical signal is given by 
\begin{equation}
	N_{qp}^{rad}=\frac{\eta_{opt}\eta_{pb}P_{rad}\Delta}{\tau_{qp}} = \frac{\eta_{opt}\eta_{pb}P_{rad}K}{\Delta N_{qp}^{read}}, 
\end{equation}
with $\eta_{opt}$ the optical efficiency and $\eta_{pb}$ the pair breaking efficiency. The total number of quasiparticles is thus given by
\begin{equation}
 	N_{qp}=N_{qp}^{read}+N_{qp}^{rad} = \frac{\eta_{opt}\eta_{pb}P_{rad}K}{\Delta\sqrt{\frac{\eta_{read}P_{read}}{K\Delta}}} + \sqrt{\frac{\eta_{read}P_{read}}{K\Delta}}.
\end{equation}
We can now derive $dN_{qp}/dP_{rad}$:
\begin{equation}
	dN_{qp}/dP_{rad} = \frac{\eta_{opt}\eta_{pb}K}{\Delta\sqrt{\frac{\eta_{read}P_{read}}{K\Delta}}}\propto P_{read}^{-1/2}. 
\end{equation}

\begin{figure}
\includegraphics[width=0.32\columnwidth]{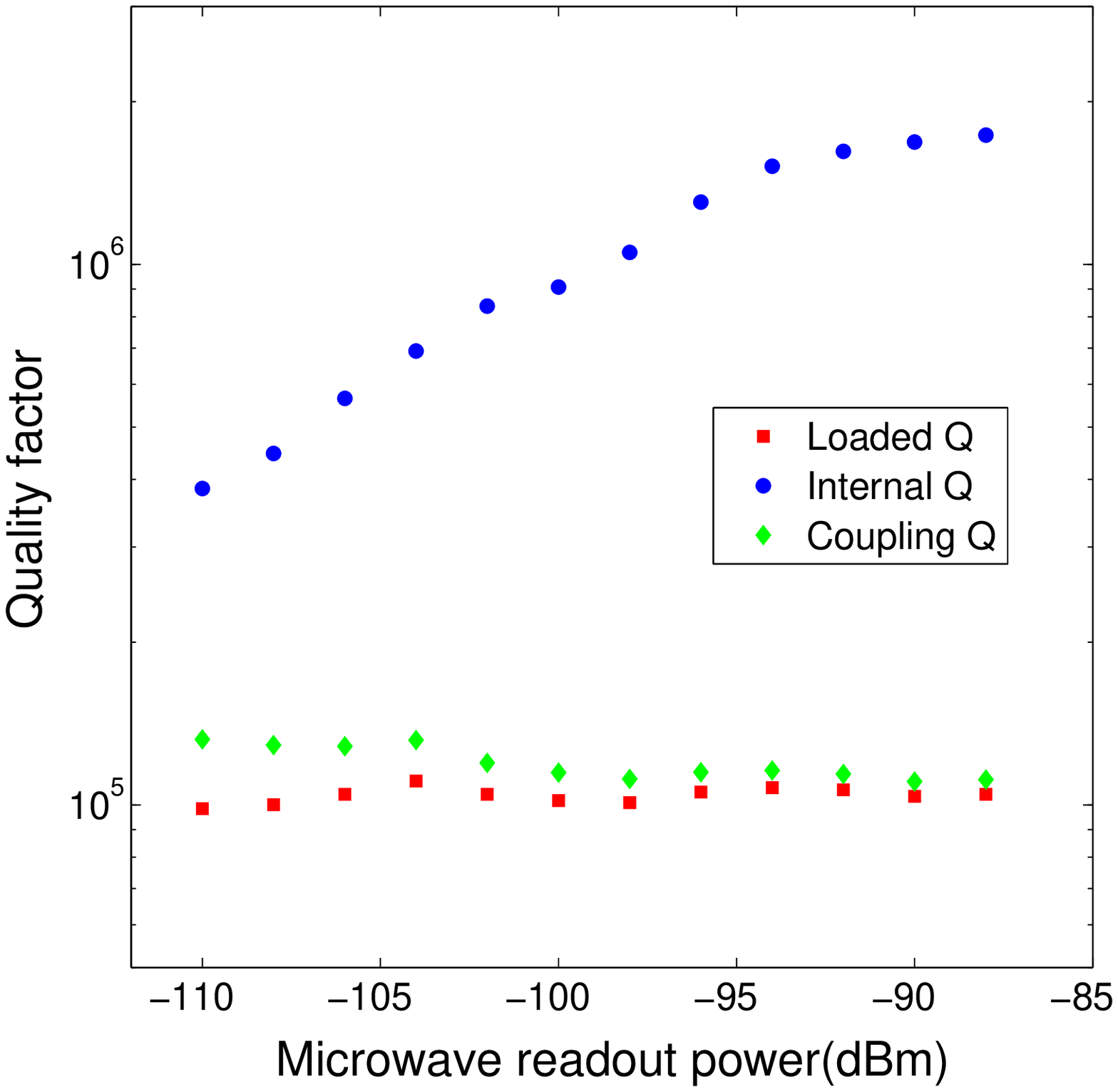}
\caption{\label{figS:QvsPread}The quality factors as determined from the microwave transmission $S_{21}$ as a function of microwave readout power at the lowest radiation power (the same radiation power as for Fig. 5 in the main article).}

\end{figure}

The resonator amplitude responsivity is now given by $dA/dP_{rad} = dA/dN_{qp}\cdot dN_{qp}/dP_{rad}$ and $dA/dN_{qp}$ is given by
\begin{equation}
	\frac{dA}{dN_{qp}} = \frac{\alpha\beta}{2}\frac{Q}{V}\frac{d\sigma_1}{|\sigma| dn_{qp}},
\end{equation}
where $\alpha$ is the kinetic inductance fraction $V$ the volume and $\beta=1+\frac{2d/\lambda}{\sinh(2d/\lambda)}\approx 2$, with $d$ the film thickness and $\lambda$ the magnetic penetration depth. The quality factor $Q$ was measured to be constant as function of readout power, as shown in Fig. \ref{figS:QvsPread} and also $|\sigma|$ is constant. $d\sigma_1/dn_{qp}$ is a slow function of effective temperature and will change only little over the measured range \cite{jgao2008cb}. Therefore we expect $dA/dP_{rad}\propto P_{read}^{-1/2}$.

In this derivation we assumed that the absorbed microwave power in the quasiparticle system $P_{abs}\propto P_{read}$. In general this depends on the details of the microwave circuit, and $P_{abs}$ and $P_{read}$ are related by \cite{jzmuidzinas2012b}
\begin{equation}\label{eq:Pdissqp}
P_{abs} = \frac{P_{read}}{2}\frac{4Q^2}{Q_iQ_c}\frac{Q_i}{Q_{i,qp}},
\end{equation}
where $Q_i$ and $Q_c$ are the internal and coupling quality factors, which are both easily measurable. $Q_{i,qp}$ is the quasiparticle quality factor, which is not known for this device, since $Q_i$ is not limited by quasiparticle dissipation. One would expect $Q_{i,qp}$ to increase for lower $P_{read}$ \cite{dgoldie2013b}, which would make the readout power dependence of the lifetime and the responsivity stronger. However, in general $Q_{i,qp}$ cannot be directly derived from $N_{qp}$ \cite{dgoldie2013b}, it depends on the shape of the driven quasiparticle distribution. We have therefore assumed for simplicity that $Q_{i,qp}$ is constant as a function of $P_{read}$. 

\section{Readout power dependence for high radiation powers}

\begin{figure}
\includegraphics[width=0.245\columnwidth]{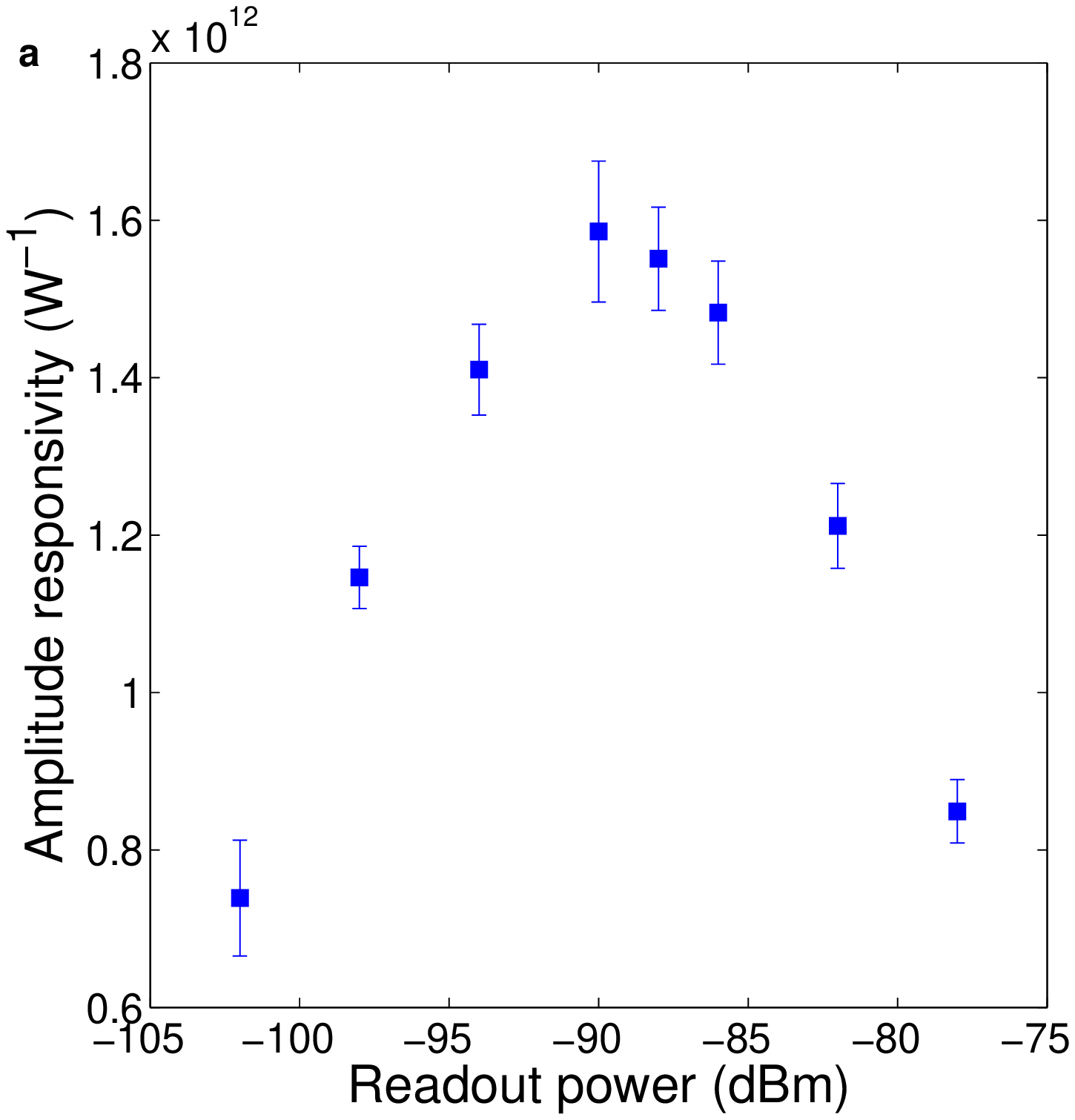}
\includegraphics[width=0.245\columnwidth]{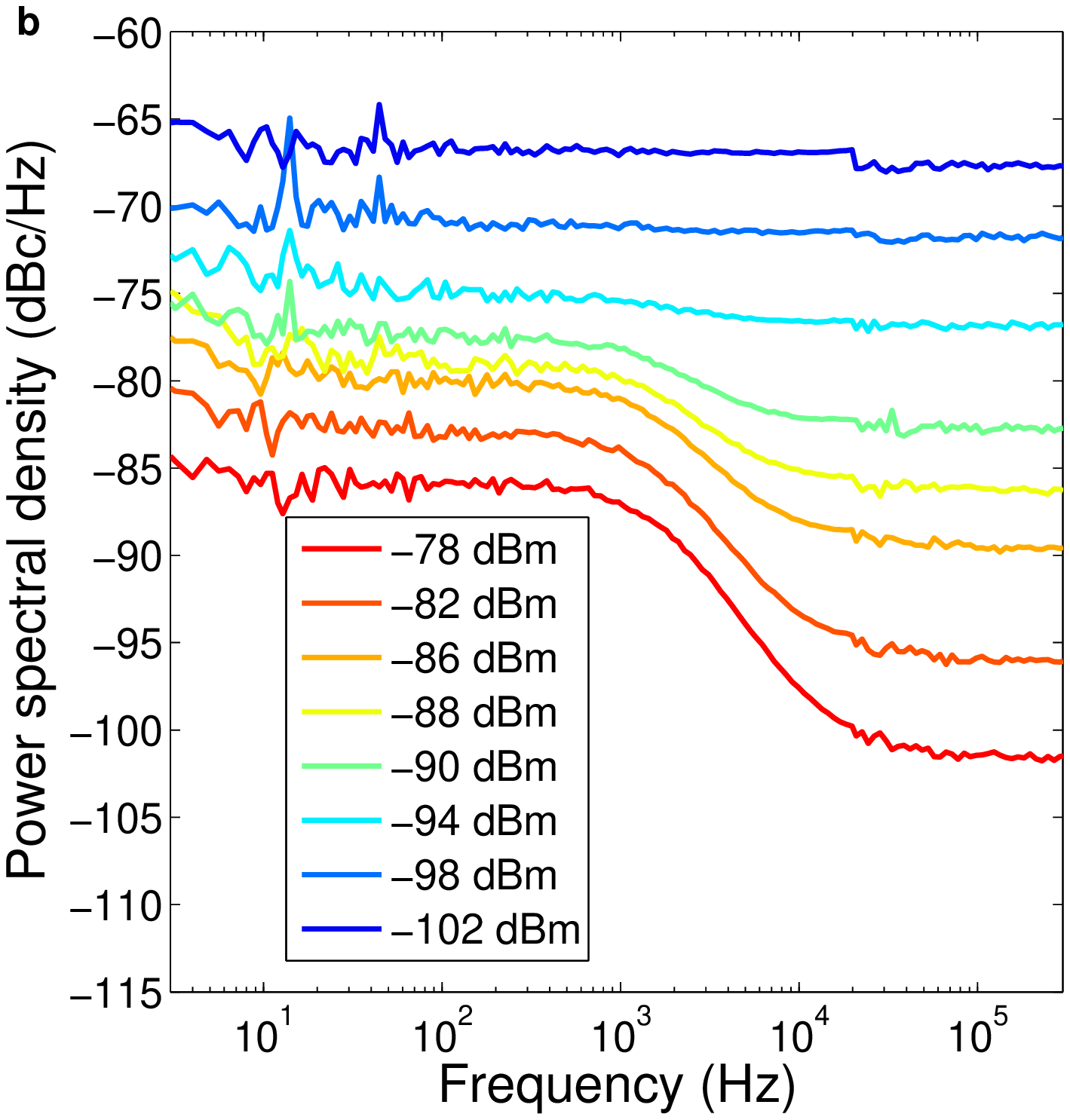}
\includegraphics[width=0.245\columnwidth]{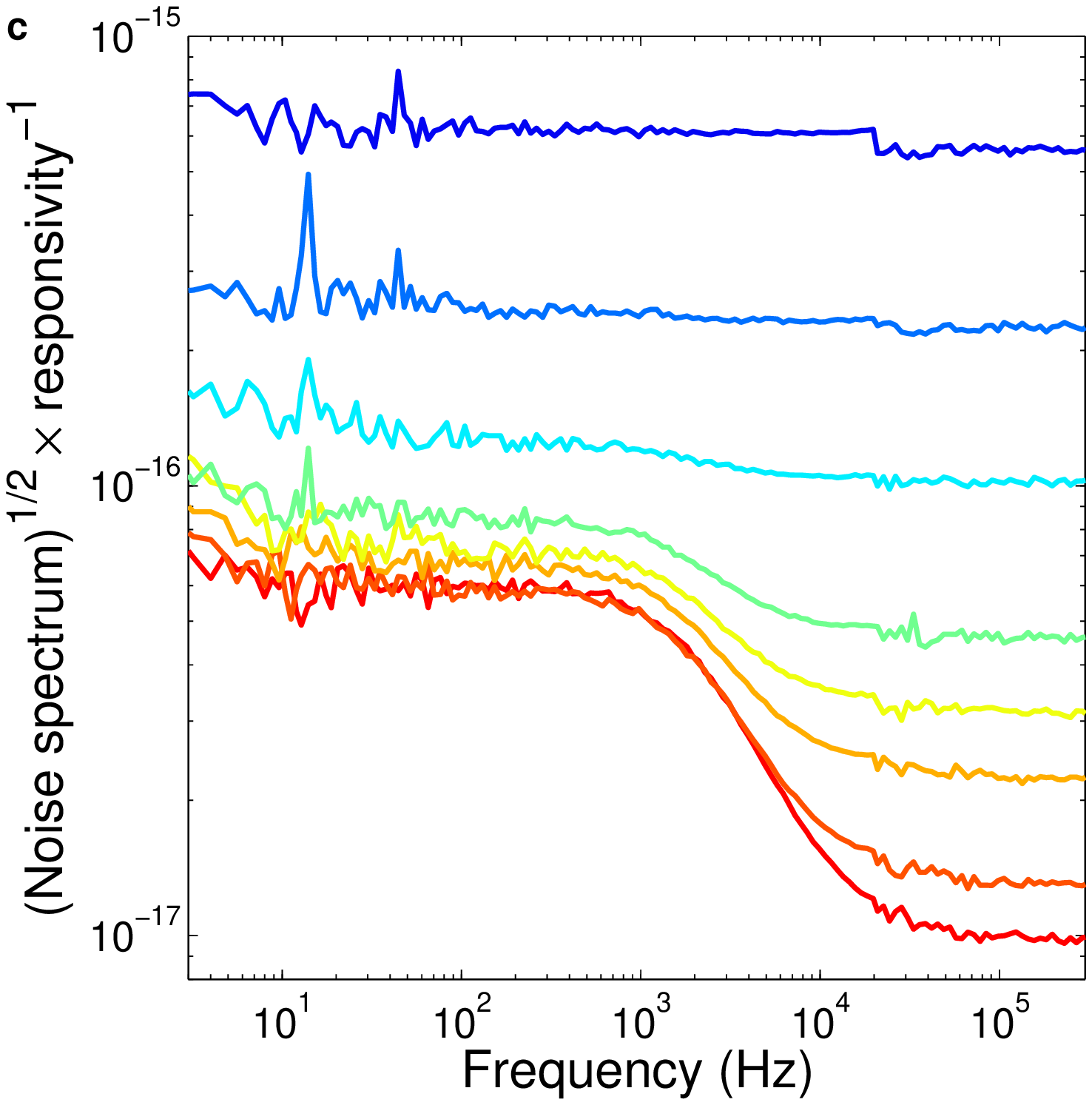}
\includegraphics[width=0.245\columnwidth]{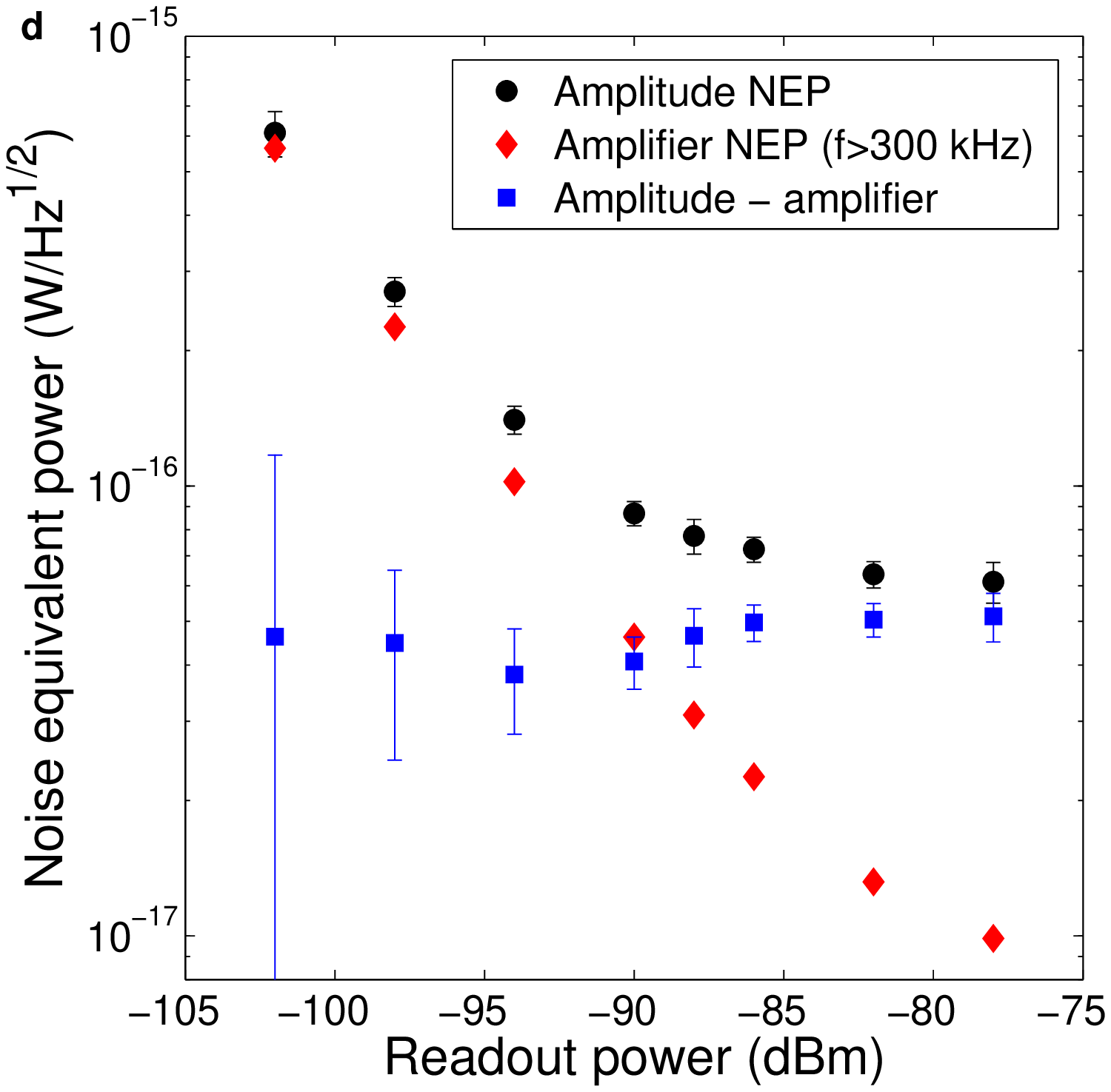}
\caption{\label{figS:NEPreadopt} \textbf{a}, The responsivity of the amplitude to radiation power $dA/dP_{rad}$ as a function of microwave readout power for the highest measured radiation power of 724 fW. \textbf{b}, The amplitude noise spectra as a function of frequency for various microwave readout powers as indicated in the legend. \textbf{c}, The optical NEP, except for the quasiparticle recombination time factor (ie $\sqrt{S_A}\cdot (dA/dP_{rad})^{-1}$), as a function of frequency for the same readout powers as in \textbf{c}. \textbf{d}, The optical NEP as a function of microwave readout power taken at the reference frequency of 20 Hz (black dots). The red diamonds show the amplifier contribution to the NEP as measured from the NEP spectra at frequencies above 300 kHz. The black squares are the NEP at 20 Hz minus the NEP above 300 kHz, thus the optical NEP with the amplifier contribution subtracted. The latter is a measure of the NEP due to photon noise only, and is therefore expected to be constant as a function of readout power. }

\end{figure}

In the main article (Fig. 5) we have discussed the influence of the readout power on the optical response for the lowest radiation power, thus in the regime where the readout power dissipation dominates the number of quasiparticles. However, also in the regime where the optical signal dominates the number of quasiparticles the optical response is readout power dependent, as is shown in Fig. \ref{figS:NEPreadopt}a for the highest measured radiation power (724 fW). Since the level of the noise spectrum depends on the responsivity (Eq. 2 of the main article), it does not surprise that the level of the photon noise roll-off also varies with the readout power, as shown in Fig. \ref{figS:NEPreadopt}b. Additionally in Fig. \ref{figS:NEPreadopt}b the amplifier noise level (the flat part at high frequencies) changes with readout power as expected. 

To see up to how far the detector sensitivity changes, we plot $\sqrt{S_A}\cdot(dA/dP_{rad})^{-1}$ (ie the NEP without the lifetime roll-off factor) in Fig. \ref{figS:NEPreadopt}c, which corrects for the responsivity of the detector and only consist of the photon noise and amplifier noise. We observe that the NEP within the photon-noise roll-off is indeed similar now for all readout powers. In Fig. \ref{figS:NEPreadopt}d we plot the NEP at the reference frequency of 20 Hz, together with an estimate of the amplifier contribution (taken at $f>$300 kHz). If we subtract the amplifier contribution we see that the leftover photon noise contribution is approximately readout power independent. We conclude that as long as the quasiparticle fluctuations are dominated by photon noise, the readout power dependence of the responsivity does not influence the detector sensitivity (NEP). In practice that means that one can use the highest possible readout power when the detector is photon noise limited to suppress the amplifier noise. Why the responsivity is readout power dependent is a complex problem that requires simulation of the influence of both radiation power and readout power absorption, a start of which has recently been made for the readout power dissipation only \cite{dgoldie2013b}.

Regarding the readout power, we make a few last remarks: 
\begin{itemize}
\item The maximum readout power before bifurcation is -88 dBm at the lowest radiation powers and increases to -78 dBm at the highest power of 724 fW.
\item In Fig. 4 and 1c of the main article we took the readout power at which the NEP is the lowest. These readout powers are (from low to high radiation power): -94 dBm, -96 dBm, -98 dBm, -92 dBm, -94 dBm, -96 dBm, -96 dBm, -88 dBm, -88 dBm, -86 dBm, -86 dBm, -78 dBm.
\item The increase in noise level upon crossing from generation-recombination noise to photon noise in Fig. 3a of the main article is only clear for a constant readout power as is shown in that figure. When different readout powers are used for different radiation powers, the changing responsivity changes the picture. We therefore did not quantitatively analyse this problem, because it requires a complex model as discussed before. The quasiparticle recombination time from the roll-off frequency is not influenced by the responsivity and therefore a more direct measure of the behaviour of the quasiparticle system.

\end{itemize}


\end{document}